\documentclass[journal]{IEEEtran}

\IEEEoverridecommandlockouts
\usepackage{cite}
\usepackage{amsmath,amssymb,amsfonts}
\usepackage{algorithmic}
\usepackage{algorithm}
\usepackage{graphicx}
\usepackage{textcomp}
\usepackage{xcolor}
\usepackage{url}
\usepackage{bm}
\usepackage{subfigure}
\usepackage{flushend}
\usepackage{CJKulem}
\usepackage{verbatim}
\usepackage{enumitem}
\usepackage{nameref}
\usepackage{ulem}

\usepackage[format=hang,labelsep=quad,margin=5pt]{caption}

\makeatletter
\newcommand{\figcaption}[1]{\def\@captype{figure}\caption{#1}}
\newcommand{\tblcaption}[1]{\def\@captype{table}\caption{#1}}

\newcommand{\Rmnum}[1]{\expandafter\@slowromancap\romannumeral #1@}
\makeatother
\makeatother
\usepackage{mathtools}

\begin{document}

\title{High-Speed Resource Allocation Algorithm Using a Coherent Ising Machine for NOMA Systems}

\author{Teppei Otsuka, Aohan~Li,~\IEEEmembership{Member,~IEEE},
Hiroki Takesue,\\
Kensuke Inaba,
Kazuyuki Aihara,
Mikio~Hasegawa,~\IEEEmembership{Member,~IEEE} 
\thanks{Teppei Otsuka, Aohan Li and Mikio~Hasegawa are with the Department of Electrical Engineering, Tokyo University of Science, Tokyo, 1258585, Japan. (E-mail: t-otsuka@haselab.ee.kagu.tus.ac.jp, aohanli@ieee.org, hasegawa@ee.kagu.tus.ac.jp).}
\thanks{Hiroki Takesue and Kensuke Inaba are with NTT Basic Research Laboratories, NTT Corporation, Atsugi, Japan. (E-mail:\{hiroki.takesue.km,  kensuke.inaba.yg\}@hco.ntt.co.jp).}
\thanks{Kazuyuki Aihara is with International Research Center for Neurointelligence, The University of Tokyo, Tokyo, Japan. (E-mail:kaihara@g.ecc.u-tokyo.ac.jp).}

\thanks{Corresponding Author: Aohan Li}
\thanks{Manuscript received XX XX, 2022; revised XX XX, 2022.}}
\maketitle

\begin{abstract}
Non-orthogonal multiple access (NOMA) technique is important for achieving a high data rate in next-generation wireless communications. A key challenge to fully utilizing the effectiveness of the NOMA technique is the optimization of the resource allocation (RA), e.g., channel and power. However, this RA optimization problem is NP-hard, and obtaining a good approximation of a solution with a low computational complexity algorithm is not easy. To overcome this problem, we propose the coherent Ising machine (CIM) based optimization method for channel allocation in NOMA systems. The CIM is an Ising system that can deliver fair approximate solutions to combinatorial optimization problems at high speed (millisecond order) by operating optimization algorithms based on mutually connected photonic neural networks. The performance of our proposed method was evaluated using a simulation model of the CIM. We compared the performance of our proposed method to simulated annealing, a conventional-NOMA pairing scheme, deep Q learning based scheme, and an exhaustive search scheme. Simulation results indicate that our proposed method is superior in terms of speed and the attained optimal solutions.

\end{abstract}

\begin{IEEEkeywords}
Non-orthogonal multiple access, resource allocation, coherent Ising machine, mutually connected neural network
\end{IEEEkeywords}

\section{Introduction}
\label{sect:introduction}

With the rapid increase in the number of mobile devices, there is an urgent need to increase the spectrum efficiency of mobile communications\cite{1}. 
A non-orthogonal multiple access (NOMA) technique has been proposed as one of the candidate technologies to meet the requirements of the next generation of wireless networks \cite{2,3}. 
Unlike conventional orthogonal multiple access (OMA) schemes, NOMA schemes can effectively increase the spectrum efficiency by introducing extra-power-domains, which enables the data of multiple different users to be multiplexed on the same channel \cite{3}.
More specifically, by applying superposition coding in the transmitter and successive interference cancellation (SIC) technology in the receiver, signals from multiple users that are multiplexed can be distinguished.

In a NOMA system, resources such as channels and power must be appropriately allocated to improve the system performance, such as the data rate, and avoid system outages such as an SIC error \cite{4,21a}.
Previous studies have shown that optimal resource allocation (RA) can increase the total achievable data rate of a system in comparison with random RA \cite{5}. 
Specifically, channel allocation methods that consider the differences in channel gain between users, and power allocation methods that balance system data rates and user fairness, are key aspects in a NOMA system design.
In addition, it is important to allocate resources to avoid the failure conditions of SIC and maximize the performance, including the total data rate, quality of service (QoS), or fairness of the NOMA system. However, joint channel and power allocation problems have been proven to be NP-hard \cite{6}, and their high complexity hinders an efficient and reliable RA, resulting in a low performance of NOMA systems.

\begin{table*}[hbtp]
\centering
\caption{The abbreviations used in this paper}
\scalebox{1.2}{
\begin{tabular}{cl||cl}
\multicolumn{1}{c|}{Abbreviation}               & \multicolumn{1}{c||}{Description}                          & \multicolumn{1}{c|}{Abbreviation}                &\multicolumn{1}{c}{Description}                     \\ \hline
\multicolumn{1}{l|}{ANN}    & Attention-based Neural Network           & \multicolumn{1}{l|}{FD-NOMA} & Full Duplex-NOMA                     \\
\multicolumn{1}{l|}{BS}     & Base Station                             & \multicolumn{1}{l|}{IPSA}    & Ideal Pairing Search Algorithm       \\
\multicolumn{1}{l|}{BPNN}   & Backpropagation Neural Network           & \multicolumn{1}{l|}{ML}      & Machine Learning                     \\
\multicolumn{1}{l|}{BHD}    & Balanced Homodyne Detection              & \multicolumn{1}{l|}{NOMA}    & Non-Orthogonal Multiple Access       \\
\multicolumn{1}{l|}{CNR}    & Carrier-to-Noise Ratio                   & \multicolumn{1}{l|}{OPA}     & Optimal Power Allocation             \\
\multicolumn{1}{l|}{CIM}    & Coherent Ising Machine                   & \multicolumn{1}{l|}{OMA}     & Orthogonal Multiple Access           \\
\multicolumn{1}{l|}{C-NOMA} & Conventional-NOMA                        & \multicolumn{1}{l|}{PSA}     & Phase Sensitive Amplifier            \\
\multicolumn{1}{l|}{DQN}    & Deep Q Learning                          & \multicolumn{1}{l|}{QoS}     & Quality of Service                   \\
\multicolumn{1}{l|}{DRL}    & Deep Reinforcement Learning              & \multicolumn{1}{l|}{RL}      & Reinforcement Learning               \\
\multicolumn{1}{l|}{DOPO}   & Degenerate Optical Parametric Oscillator & \multicolumn{1}{l|}{RA}      & Resource Allocation                  \\
\multicolumn{1}{l|}{D2D}    & Device-to-Device                         & \multicolumn{1}{l|}{SA}      & Simulated Annealing                  \\
\multicolumn{1}{l|}{ES}     & Exhaustive Search                        & \multicolumn{1}{l|}{SIC}     & Successive Interference Cancellation \\
\multicolumn{1}{l|}{FPGA}   & Field Programmable Gate Array            & \multicolumn{1}{l|}{WLAN}    & Wireless LAN                        
\end{tabular}
}
\end{table*}

To solve the NP-hard RA optimization  problems in NOMA systems, many approaches such as machine learning (ML) and heuristic approaches have been proposed  \cite{7,20,21,17,18,19,21a}.
Optimization using ML has the advantage of a rapid RA in large-scale problems by using trained models that have been learned in advance. However, a trained model may no longer be usable and must be retrained when the system environment changes. 
This has a disadvantage that the RA may not be sufficiently fast when the communication environments change.
By contrast, optimization using heuristic methods is faster owing to a lower computational complexity; however, a fundamental trade-off emerges between the solution accuracy and computational complexity.
By contrast, hardware-based search algorithms, such as quantum computers and Ising machines, have been proposed and are expected to solve various optimization problems at high speeds\cite{9,10,11,12,12a}.
As described in \cite{34}, quantum technologies are emerging technologies that are expected to be deployed in 6G networks around 2030, although it is still unclear when a practical quantum computer will be available.
D-wave\cite{9} is a quantum annealing machine that has already been commercialized and is available for general users. The coherent Ising machine (CIM) \cite{10,11,12,12a} is an  optimization machine based on laser networks, which can deliver fair approximate solutions to combinatorial optimization problems at high speed on the order of milliseconds.
Many NP-complete and NP-hard problems can be  transformed into ground-state search problems of the Ising model \cite{13a}.
Here, the CIM and D-Wave are Ising machines that can rapidly obtain a state near the ground state of the Ising Hamiltonian which represents a good solution to the optimization problem.
As the main difference between the D-Wave and CIM, D-Wave reproduces a coupling of magnetic spins with a chimera topology having sparse connections, whereas in CIM, the Ising network of optical pulses generated by a laser oscillator is fully coupled\cite{10,11}.

A previous study \cite{12} demonstrated that the CIM is more effective than D-Wave for large-scale large-density problems, and it will be more suitable for the next-generation large-scale wireless communications. In \cite{12a}, the authors showed that 100,000 fully connected spins can be implemented using the CIM. In previous studies \cite{13,14,24,25}, various wireless communication systems are optimized using CIMs. A CIM-based optimization approach has been applied to solve the NP-hard optimization problems in wireless communications. In \cite{16}, it is shown that the optimal solutions obtained can be empirically scaled as $O(1)$ using the CIM. It is therefore possible to achieve a real-time optimization of RA in a large-scale NOMA system using the CIM. In this paper, we propose a RA method for NOMA systems using the CIM. The main contributions of this paper can be summarized as follows.

\begin{itemize}
\item We propose a high-speed RA method using a CIM for NOMA systems, especially for the channel assignment. The proposed method is a hardware-based method with superior speed (milliseconds-order) compared to other existing methods.
\item To the best of our knowledge, previous works related to RA using the CIM in wireless communication assumed that only one user could access each channel simultaneously. In contrast to previous studies, this study focuses on the RA optimization problem using the CIM for NOMA systems where multiple users can be multiplexed in the same channel. Hence, novel solutions are necessary to solve such problems which are different from those in our previous work. For instance, dummy users need to be introduced and the constraints are different, which leads to different strength of the interaction between spins and that of the external magnetic field on each spin. In this study, we formulated the RA combinatorial optimization problem carefully to meet the characteristics of the RA in the NOMA system and the feasibility of solving it using the CIM.
\item To solve the RA optimization problem in NOMA systems using the CIM, it is necessary to derive the parameters of the Ising model and introduce them into the CIM. Therefore, we first transformed the formulated RA combinational problem into a ground state search problem of the Ising Hamiltonian. Then, we derived the parameters of the Ising model that needed to be introduced into the CIM to obtain the solutions, i.e., the interaction between spins and the external magnetic field.
\item To verify the effectiveness of the proposed method, we first evaluated the convergence of the proposed method. Simulation results demonstrated the convergence of the proposed methods. Then, we evaluated the performance of the proposed method in the total data rate and compared it with that of other methods including simulated annealing (SA), exhaustive search (ES), the conventional-NOMA (C-NOMA), deep Q learning (DQN), and random methods. The simulation results indicate that our proposed approach can achieve higher data rates than other methods. The computation time of the proposed, SA, DQN and ES methods was evaluated, which demonstrates that the proposed method can obtain better performance at faster speed.

\end{itemize}

The rest of this paper is organized as follows. Section \Rmnum{2} reviews previous related studies. Section \Rmnum{3} introduces the system model and problem formulation. Section \Rmnum{4} introduces the principle of the CIM in solving the optimization problems.
Section \Rmnum{5} presents the proposed CIM-based RA method for NOMA systems.
Section \Rmnum{6} demonstrates the simulation results.
Section \Rmnum{7} describes future work.
Section \Rmnum{8} provides some concluding remarks regarding this research. 
The abbreviations used in this paper are listed in Table I.

\section{Related Work}
\label{sect:systemmodel}
In this section, we review the related studies on RA. We first review the heuristic RA approaches and ML-based RA approaches for NOMA systems. We then review the hardware-based RA, particularly for the existing studies on CIM-based RA in wireless communications that are not limited to NOMA RA problems.

\subsection{Heuristic RA Approaches}
The high complexity of the RA problem in NOMA systems can be reduced using heuristic methods by dividing the problem into sub-problems \cite{7,17,18,19,21a}.
In \cite{21a}, the authors propose low complexity and efficient method for RA in a downlink NOMA system. In this study, user fairness, data rate, and energy efficiency are optimized. The RA problem is divided into the sub-problems of power allocation and channel allocation. The power allocation solution is derived from a theoretical analysis. The channel allocation is carried out based on the derived power allocation solution.
In \cite{17}, the RA in full duplex-NOMA (FD-NOMA) systems is studied. In this study, the RA problem is successfully solved with low computational complexity by dividing the channel and power problems based on the block coordinate descent method.
In \cite{18}, an algorithm for downlink NOMA systems that analytically examines the optimal power allocation (OPA) and ideal pairing search algorithm (IPSA) in terms of fairness was proposed.
Simulation results indicate that the computational complexity is significantly reduced in comparison to the ES.
In \cite{19}, NOMA-based device-to-device (D2D) communications are considered. It is shown that the communication performance can be improved to a significant extent using the NOMA technique. The RA solution is obtained with low computational complexity by applying an optimal power allocation after a channel allocation algorithm that assigns sub-carriers to the D2D groups.
In \cite{7}, the RA in a NOMA system is optimized using the SA, which is a meta-heuristic algorithm. In this study, a joint optimization of the channel allocation, sub-channel power allocation, and inter-user power allocation is considered.
Simulation results indicate that a sufficiently reliable solution in terms of system throughput can be obtained with low computational complexity in comparison to an ES.

As discussed above, we can see that the heuristic RA approaches can obtain a solution with low computational complexity. However, there is a fundamental trade-off between the performance of the solution and the computational complexity, and there have been no previous studies that can quickly optimize on the order of milliseconds.

\subsection{ML Based RA Approaches}
With the development of ML applications in wireless communications, ML solutions for RA optimization problems in NOMA systems have been proposed \cite{4,20,21}.
In \cite{4}, deep reinforcement learning (DRL) based RA scheme is proposed for NOMA systems. Here, the framework of the DRL method is designed based on an attention-based neural network (ANN). 
Specifically, the joint optimization problem of channel and power allocation is formulated, and the channel allocation is learned according to the policy learned from the ANN.
The simulation results indicate that the proposed method can achieve the same performance for six users with lower computational complexity in comparison to the ES.
In \cite{20}, a DQN is adopted to achieve a joint RA for a large number of users, where reinforcement learning (RL) frames are used for the joint optimization of clustering, power allocation, and beamforming.
Specifically, the clustering is adjusted using a DQN, and a backpropagation neural network (BPNN) is designed to learn the power coefficients for each cluster. Simulation results indicate that the performance of the proposed method is close to that of the ES. 
In \cite{21}, a method based on DRL is proposed to assign users to time-varying channels. Recently,
DRL has been expected to be applied as a solution for such real-time dynamic decision-making problems in practical environments where the conventional RA algorithms are difficult to be applied. 
A complexity analysis and simulation results show that the proposed approach provides a superior performance with lower computational complexity in comparison to conventional optimization algorithms. 

Therefore, ML-based RA methods can obtain near-optimal solutions with lower computational complexity than the ES. However, to obtain near-optimal solutions, a proper training time is required.

\subsection{Hardware-based RA}
Optimization problems such as  channel and power allocation are important not only in NOMA systems but also in a variety of other wireless communication systems. Optimization problems in wireless communication systems have recently become increasingly complex, and the demand for efficient solutions to meet the requirements of next-generation wireless communications has increased \cite{22}.
Hardware-based approaches, such as quantum computers and Ising machines, have been expected to be potential methods for solving optimization problems at high speed \cite{9,10,11,12}. 
In previous studies, several RA optimization problems in wireless communications have been studied using the CIM \cite{13,14,24,25}.
In \cite{14}, the CIM is applied to RA problems in wireless communication systems such as IEEE802.11 wireless LAN systems and D2D communications. In addition, it has been shown that the CIM can approximately solve these optimization problems within the order of milliseconds.
In \cite{24}, the CIM is applied to the optimization of scheduling problems for distributed antenna systems.
In \cite{25}, the CIM is applied to the optimization problem of channel assignment in dense wireless LAN (WLAN) systems.
In this study, the maximization of the system throughput in high-density WLAN systems using the CIM is considered.
The performance of the proposed CIM-based channel assignment approach in comparison to other optimization approaches such as the SA and greedy algorithms is evaluated.
Simulation results indicate that the CIM can obtain the optimal solution well with constant computational time, whereas other optimization algorithms increase the computational time and decrease the accuracy of the obtained solution as the number of users in the system increases.

In these studies, the CIM is assumed to work either as a base station (BS) or as a controller in the centralized systems. The ability of the CIM to operate at room temperature \cite{22a} and its simple components make it suitable for integration into many wireless transmission systems.
Hence, the CIM is significantly more suitable for large-scale wireless networks
if the size of the problem is up to the number of spins that can be achieved with the CIM.

In summary, the CIM may obtain optimal solutions to combinatorial optimization problems without the need for complex computer calculations. The principle of searching for optimal solutions using the CIM is based on the spontaneous energy reduction in physical phenomena. Existing research has demonstrated the higher speed of a CIM-based optimization. On the contrary, DRL or other ML algorithms require training samples to obtain the optimal solution, and trial and error to increase performance.
In a real communication environment, such as a vehicle network, the environment changes frequently, and real-time optimization may not be possible for the ML algorithms because retraining is needed when the environment changes. In \cite{4,21}, the complexity due to training in the ML algorithm was studied. In contrast to the complexity of training presented in these papers, the proposed CIM-based RA method can always solve the optimization problem at constant high speed, i.e., millisecond order \cite{16}. Therefore, real-time RA can be achieved using our proposed method, which may improve the spectrum efficiency under dynamic changing environments.

\section{System Model and Problem Formulation}
\label{sect:systemmodel}

\begin{figure}[!t]
    \centering
    \includegraphics[width=90mm]{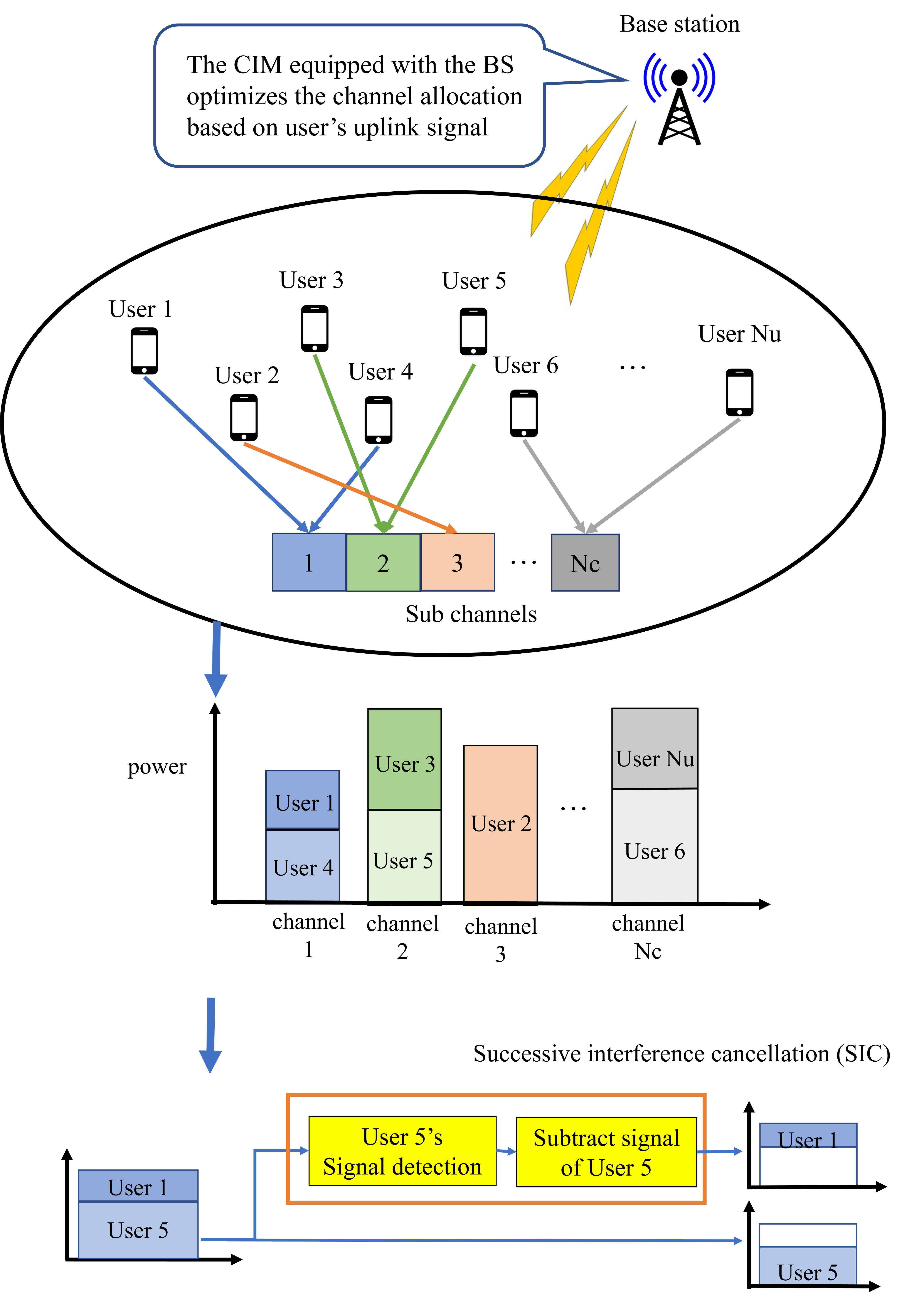}
    \caption{System model.}
    \label{fig:systemmodel}
\end{figure}

We consider a downlink NOMA system where the BS transmits data to multiple users, as shown in Fig. \ref{fig:systemmodel}.

Both the BS and users have a single antenna.
At the BS, the transmitter sends signals with multiplexing the data of different users on the same channel.
Assume that the signal to each user is multiplexed in one channel at the same time.
At the receiver side, the SIC technology is used to restore the signal of each user.
The BS is equipped with the CIM. Once the required communication information is gathered in the BS side, the channel will be allocated by the CIM.
Let the set of users in the NOMA system be ${\bf U}=\{1,2,\ldots,u,\ldots,N_u\}$ and the set of sub-channels be ${\bf C}=\{1,2,\ldots,c,\ldots,N_c\}$, where $N_u$ and $N_c$ are the numbers of the users and channels in the NOMA system. Note that, $N_u>2$.
Assume that the total bandwidth is $B$ while the bandwidth is divided into $N_c$ sub-channels equally.
Hence, the bandwidth of the $c$th channel is $B_c=B/N_c$.
The multiplexed signal on the $c$th channel can be expressed as follows:
\begin{equation}
\label{eq1}
x_c=\sum_{i=1}^{Mc}\sqrt{P_i^c}b_i ,
\end{equation}
where $M_c$ is the number of users multiplexed on the $c$th channel, $P_i^c$ is the power allocated to the $i$th user transmitted on the $c$th channel, and $b_i$ is the transmission symbols of the $i$th user.
At the receiver side, the multiplexed signal of the $u$th user in the $c$th channel can be expressed as follows:
\begin{equation}
\label{eq2}
y_u^c=\sqrt{P_u^c}h_u^cb_u+\sum_{{\substack{i=1 \\ i\neq u}}}^{M_c}{\sqrt{P_i^c}h_i^cb_i}+z_u^c ,
\end{equation}
where $z^c_u$ is additive white Gaussian noise with zero mean and variance $\sigma^2_{z_c}$.
$h^c_u$ is the channel response between the BS and the $u$th user, which can be calculated as follows:
\begin{equation}
\label{eq3_1}
\left|h_u^c\right|^2=\left(g_u^c\right)^2d_u^{-\alpha} ,
\end{equation}
where $g^c_u$ is a coefficient that follows a Rayleigh distribution, $d_u$ is the distance between the BS and the $u$th user, and $\alpha$ is the path loss coefficient. The carrier-to-noise ratio (CNR) in the $c$th channel of the $u$th user is defined as follows:
\begin{equation}
\label{eq3_2}
\Gamma_u^c=\frac{\left|h_u^c\right|^2}{\sigma_{z_c}^2} .
\end{equation}
Herein, we assume that the CNR of each user assigned to a channel $c$ is $\Gamma_1^c>\ldots>\Gamma_u^c>\ldots>\Gamma_{M_c}^c$.
In the NOMA protocol, higher power is allocated to the users with lower CNR \cite{21a,21c}, i.e., $P_1^c<\ldots<P_u^c<\ldots<P_{M_c}^c$. In addition, SIC is performed at the receiver side, which successively removes the signal of the user with the higher power. Therefore, the signal of the $u$th user can be decoded by removing the signal of the $i$th user when $P_i^c>P_u^c$. The signal of the $u$th user is considered as the interference of the $i$th user at that time. Based on the principle discussed above, when the SIC is fully working in the power domain NOMA, i.e., when there is a sufficient difference in gain between users that are multiplexed in the same channel, the signal-to-interference-plus-noise ratio of the $u$th user can be calculated as follows:

\begin{equation}
\label{eq4}
\gamma_u^c=\frac{P_u^c\Gamma_u^c}{1+\sum_{i=1}^{u-1}P_i^c\Gamma_u^c} .
\end{equation}
The data rate achievable by the $u$th user in the $c$th channel can then be calculated as follows:
\begin{equation}
\label{eq5}
R_u^c=B_c\log_2{\left(1+\frac{P_u^c\Gamma_u^c}{1+\sum_{i=1}^{u-1}P_i^c\Gamma_u^c}\right)} .
\end{equation}
Assuming that the total transmission power from the BS is $P_T$, the relation between the power coefficient $P_u^c$ and total transmission power $ P_T$ can then be expressed as follows:
\begin{equation}
    \sum_{u=1}^{N_u}\sum_{c=1}^{N_c}P_u^c\leq P_T .
\end{equation}
As the number of users assigned to the same channel increases, the complexity of SIC decoding at the receiver side increases, which leads to high complexity of the power allocation scheme to efficiently avoid SIC errors \cite{21a,21}.
SIC and power allocation are needed to be carefully considered if we assume that more than 2 NOMA access the same channel simultaneously, which is beyond the scope of this paper. The contribution of this paper is to achieve fast RA using the CIM for NOMA systems. Therefore, we focus on introducing how to use the CIM to solve the RA problem in NOMA.
In other words, we consider the scenario where $M_c=2$ for $\forall c\in \bf{C}$.
Thus, the data rate that can be achieved by any user in any channel can be expressed
 \begin{eqnarray}
  \label{eq6}
    R_{NOMA}^{ijk}= \begin{dcases}
    B_c\log_2{\left(1+P^j_{ik}\Gamma_i^j\right)},\ &\mathrm{if}\ \Gamma_i^j>\Gamma_k^j ,\\
    B_c\log_2{\left(1+\frac{P^j_{ik}\Gamma_i^j}{1+P^j_{ki}\Gamma_i^j}\right)},\  &\mathrm{otherwise} ,
  \end{dcases}
\end{eqnarray}
where for $i\in {\bf U},\ k\in {\bf U} , \text{and} \ j\in {\bf C} ,\nonumber$  $R_{NOMA}^{ijk}$ is the data rate achieved by the $i$th user when multiplexed with the $k$th user in the $j$th channel, and $P^j_{ik}$ is the transmission power allocated to the $i$th user when the $i$th user is multiplexed into the $j$th channel with the $k$th user.
In addition, when there is only one user $i$ is assigned to channel $j$, the data rate of the $i$th user can be expressed as follows:
\begin{equation}
    R^{ij}_{OMA}=B_c\log_2{\left(1+P^j_{i}\Gamma^j_i\right)},
\end{equation}
where $P_i^j$ is the transmit power allocated to the $i$th user on the $j$th channel. Here, the expression of the relationship between the number of users and that of channels below must be stratified to transform the formulated problem into an Ising problem. That is, $N_u=2N_c$. However, we assume that up to 2 users can be multiplexed to the same channel, which means that one or no users may be assigned to the channel. The transmissions will be OMA when only one user is assigned to the channel. Therefore, in addition to the user set $\bf U$, we define the dummy user set as ${\bf D}=\{1,\ldots,N_d\}$. If the user from set $\bf U$ is multiplexed with the user from set $\bf D$, OMA transmission will be carried out by the user from set $\bf U$, which means that the user in the set $\bf D$ is not a really existing user in the real world. As described above, we can understand that the total number of users and that of the dummy users is twice the number of channels i.e., $N_d+N_u=2N_c$. Hence, the throughput achieved by users in the NOMA system can be redefined as follows:
\begin{eqnarray}
  \label{eq6a}
    R^j_{ik}= \begin{dcases}
    R_{NOMA}^{ijk},\ \ \ \ &\mathrm{if}\ \ i,k\in \mathrm{\bf{U}} ,\\
    R^{ij}_{OMA},\ \ \ \ &\mathrm{if}\ \ i\in \mathrm{\bf{U}} ,\ k\in \mathrm{\bf{D}} ,\\
    0,\ \ \ \ &\mathrm{if}\ \ i\in \mathrm{\bf{D}},\ \forall k.\\
  \end{dcases}
\end{eqnarray}

In this paper, our goal is to maximize the total data rate of the NOMA system. The objective function of this paper is expressed as follows:
\begin{align}
\label{eq8} \genfrac{}{}{0pt}{}{\text{max}}{x}
    &\sum_{i=1}^{N_u+N_d}\sum_{j=1}^{N_c}\sum_{{\substack{k=1 \\ k\neq i}}}^{N_u+N_d}\left(R^j_{ik}+R^j_{ki}\right)x_{ij}x_{kj}\\
    &{\it s.t.}\;\;\ R^j_{ik}\geq\;(R_{ij})_{min},\ \ \mathrm{for}\ \ \forall k ,\tag{\ref{eq8}.a}\\
    &\ \ \ \ \ \ \sum_{i=1}^{N_u+N_d}\sum_{j=1}^{N_c}\left(P_{ik}^j+P_{ki}^j\right)x_{ij}x_{kj}\leq P_T\tag{\ref{eq8}.b},\\
    &\ \ \ \ \ \ \ \ \sum_{j=1}^{N_c}x_{ij}=1,\ \ \mathrm{for}\ \ \forall i ,\tag{\ref{eq8}.c}\\
    &\ \ \ \ \ \ \sum_{i=1}^{N_u+N_d}x_{ij}=2,\ \ \mathrm{for}\ \ \forall j ,\tag{\ref{eq8}.d}
\end{align}
where $x_{ij}$ represents the channel assignment variable, which can be expressed
\begin{eqnarray}
 \label{eq9}
    x_{ij}= \begin{cases}
    1,\ \ \ \ \mathrm{if \ the}\ i\mathrm{th}\ \mathrm{user}\ \mathrm{uses}\ \mathrm{the}\ j \mathrm{th}\ \mathrm{channel}, \\
    0,\ \ \ \ \ \mathrm{otherwise}.
  \end{cases}
\end{eqnarray}
Constraint (\ref{eq8}.a) indicates the minimum data rate that the user should achieve. In addition, $(R_{ij})_{min}$ in constraint (\ref{eq8}.a) is the QoS constraint, which represents the minimum data rate that the $i$th user using the $j$th channel must achieve.
Constraint (\ref{eq8}.b) indicates the power constraint, i.e., the total power allocated to the users should not exceed the total transmission power from the BS. Constraints (\ref{eq8}.c) and (\ref{eq8}.d) indicate that one user, and up to two users, can be multiplexed into a single channel, respectively.

In this paper, power allocation is optimized using the method in \cite{21a}, whereas channel allocation is optimized using the CIM.
Specifically, we maximize Eq. (\ref{eq8}) while satisfying constraints (\ref{eq8}.c) and (\ref{eq8}.d) using the CIM, and the power factor considering constraints (\ref{eq8}.a) and (\ref{eq8}.b) is then optimized using the method described in \cite{21a}. 
The method of power allocation in \cite{21a} can maximize the total data rate while satisfying the QoS constraints and total power constraint, i.e., constraints (\ref{eq8}.a) and (\ref{eq8}.b) can be satisfied by allocating power using the method described in \cite{21a}.
Here, we assume that $\Gamma_i^j>\Gamma_k^j$ and $i,k\in \mathrm{\bf{U}}$. Then, the optimal $P_{ik}^j$ and $P_{ki}^j$ in Eq. (\ref{eq8}) can be derived as

\begin{equation}
\label{eq11}
\begin{split}
&P_{ik}^j=\frac{\Gamma_k^jq_j-A_{k}^j+1}{A_{k}^j\Gamma^j_{k}},\\
&P_{ki}^j=q_j-P_{ik}^j,
\end{split}
\end{equation}
where $A_{k}^j=2^{\frac{(R_{kj})^{min}}{B_c}}>2$, 
and $q_j$ is the transmission power from the BS allocated to the $j$th channel.
By using this equation, the signals can be successfully decoded as avoiding SIC failure conditions is considered. Note that if user $i \in {\bf U}$ is multiplexed with dummy user $k \in {\bf D}$ for OMA communication, $P_{ik}^j$  is set as a constant $q_j$ because the power allocated to the dummy user is zero.
The value of $q_j$ maximizing the total data rate can be obtained using the following equations:

\begin{equation}
\label{eq12}
\begin{split}
&q_j=\left[\frac{B_c}{\Lambda} - \frac{A_{k}^j}{\Gamma_i^j} + \frac{A_{k}^j}{\Gamma_k^j} - \frac{1}{\Gamma_k^j} \right]^\infty_{\gamma_j} ,\\
&\gamma_c=\frac{A^j_k\left(A^j_i-1\right)}{\Gamma_i^j} + \frac{A^j_k-1}{\Gamma_k^j} ,
\end{split}
\end{equation}
where $\Lambda$ needs to be selected to satisfy $\sum^{N_c}_{j=1}q_j=P_T$.
$\left[\bullet\right]$ is a well-known water-filling form \cite{21a}. It means the power allocation solution for the formulated combinatorial optimization problem, i.e., Eq. (\ref{eq8}) is uniquely determined if $q_j$ is in the range $\left[\gamma_j,+\infty\right)$.
In the following, we introduce the optimal channel allocation using the CIM.

\section{Coherent Ising Machine}
\label{sect:simulation}
In this section, we first introduce the Ising Hamiltonian.  Then, we present the measurement feedback CIM, which is a measurement-feedback type CIM that has been proposed to easily increase the number of couplings between spins to solve larger-scale problems. Finally, we investigate the operation process of the measurement feedback CIM. 

\subsection{Ising Hamiltonian}
The CIM is an Ising-type machine that can artificially reproduce an Ising model, a physical model of magnetic spins.
The Ising model consists of two states of magnetic spins: upward and downward.
Spins are mutually coupled and subject to interactions from other spins and effects from external magnetic fields.
We consider Ising spins with an $N\times M$ two-dimensional structure.
Here, the energy function of the Ising model, i.e., the Ising Hamiltonian is expressed as follows:

\begin{equation}
\label{eq13}
E(\sigma)=-\frac{1}{2}\sum_{i=1}^{N}\sum_{j=1}^{M}\sum_{k=1}^{N}\sum_{l=1}^{M}{J_{ijkl}\sigma_{ij}\sigma_{kl}}+\sum_{i=1}^{N}\sum_{j=1}^{M}{\lambda_{ij}\sigma_{ij}} ,
\end{equation}
where $\sigma_{ij}\in\{-1,+1\}$ is the spin direction of the $(i,j)$th spin, $J_{ijkl}$ is the strength of the interaction between the $(i,j)$th and $(k,l)$th spins, and $\lambda_{ij}$ is the strength of the external magnetic field on the $(i,j)$th spin.
Here, the CIM is the machine that can obtain the ground state of the Ising Hamiltonian at a high speed.
In other words, by setting $J$ and $\lambda$ to correspond to the optimal solution of the optimization problem to the minimum of Eq. (\ref{eq13}),
we can approximately obtain a fast optimal solution using the CIM.

\begin{figure}[!t]
    \centering
    \includegraphics[width=80mm, clip]{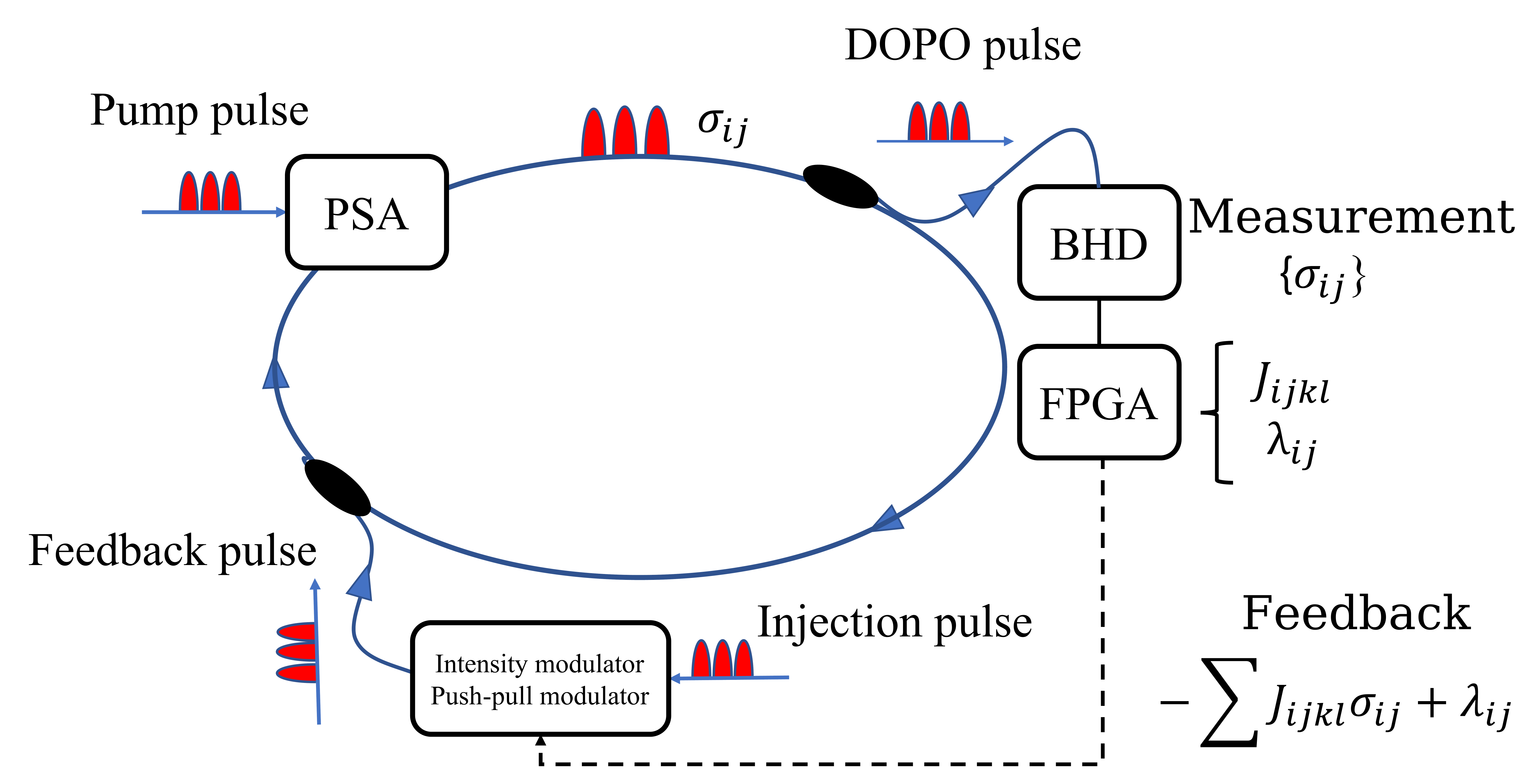}
    \caption{Coherent Ising Machine \cite{10,11}.}
    \label{fig:Coherent Ising Machine}
\end{figure}

\subsection{Measurement Feedback Coherent Ising Machine}
Initially, the CIM was a laser network consisting of one master laser and multiple slave lasers\cite{CIM_prin}. Currently, a measurement-feedback type CIM has been proposed to easily increase the number of couplings between spins to solve larger scale problems\cite{10,11}.
A schematic diagram of the measurement feedback-type CIM is shown in Fig. \ref{fig:Coherent Ising Machine}.
As shown in Fig. \ref{fig:Coherent Ising Machine}, parameters of the Ising model $J_{ijkl}$ and the $\lambda_{ij}$, parameters of the Ising model are pre-configured in a field programmable gate array (FPGA) module.
The FPGA module calculates the feedback value using the pre-configured parameters and feeds it back to the original pulse.
Therefore, interactions between spins can be programmatically reproduced, which makes it possible to fully connect spins.
In the measurement-feedback CIM, the phase of the degenerate optical parametric oscillator (DOPO) pulse circulating on an optical fiber represents Ising spins $\sigma_{ij}$.
In \cite{12a}, it is shown that 100,000 DOPO pulses can be realized over 5 km of polarization-maintaining optical fiber and reach the reference score of the MAX-CUT problem for 100,000 nodes within a time of 593 µs.

\subsection{Operation Process of the Measurement Feedback CIM}
The CIM is an Ising machine that can rapidly obtain the ground state of the Ising Hamiltonian by getting the combinations of the Ising spin directions. The set of the Ising spin directions that minimize the Ising Hamiltonian corresponds to the solution of the combinational optimization problem. The operation process of the CIM can be summarized as follows. As shown in Fig. 2, the initial pulses are first generated by a phase sensitive amplifier (PSA) on the optical fiber. The initial pulses are repeatedly amplified by PSA with a pump pulse each cycle and become a DOPO pulse. The phase of the DOPO pulse circulating on the fiber during oscillation is used to reproduce the Ising spin on the optical fiber in the CIM. Here, gradually increasing the pump pulse amplitude until the threshold is exceeded; the DOPO pulse will oscillate and take on a $\pi$ or 0 phase representing the orientation of the Ising spin $\sigma_{ij}$, where $\sigma_{ij}\in\{-1,+1\}$. 0 phase represents the up spin, i.e., $\sigma_{ij}=1$. Meanwhile, the $\pi$ phase represents the down spin, i.e., $\sigma_{ij}=-1$ While the DOPO pulse is being amplified, a part of the DOPO pulse on the fiber is separated by a coupler, and the in-phase amplitude is measured through a balanced homodyne detection (BHD). The measured in-phase amplitude $\sigma_{ij}$ is input into the FPGA module. A feedback signal $-\sum{J_{ijkl}\sigma_{ij}}+\lambda_{ij}$ is calculated using the mutual coupling between Ising spins $J_{ijkl}$ and external magnetic field $\lambda_{ij}$. Finally, the feedback signal is injected into the original DOPO pulse as a feedback pulse through the intensity modulator and push-pull modulator. By repeating the operation process of the CIM in this manner, we can eventually obtain a combination of $\sigma_{ij}$ that minimizes Eq.  (\ref{eq13}), which corresponds to the optimal solution to the optimization problem.

From the above, to obtain the optimal solution of optimization problem using the CIM, we need to transform the objective function into the form of an Ising Hamiltonian, and then derive the corresponding $J_{ijkl}$ and $\lambda_{ij}$ of the objective function. Then, by pre-configuring the derived these parameters into the FPGA module of the CIM, we can search for the ground state of the Ising Hamiltonian, which corresponds to the optimal solution to the optimization problem.  
In the next section, we introduce the problem transformation and the derivation of the corresponding $J_{ijkl}$ and $\lambda_{ij}$ for our formulated objective function in Eq. (\ref{eq8}).

\section{Applying the CIM to the RA Problem in NOMA}
\label{sect:applyingcim}

To solve the combinatorial optimization problem in the CIM, we must set $J_{ijkl}$ and $\lambda_{ij}$ in the FPGA module of the CIM.
However, it is difficult to transform the formulated objective function, i.e., Eq. (\ref{eq8}), directly into an Ising Hamiltonian using the $\{-1,+1\}$ variables corresponding to the value of the Ising spin of the CIM.
We therefore first express Eq. (\ref{eq8}) using binary variables of $\{0,1\}$, and then transform it into the form of an Ising Hamiltonian using the $\{-1,+1\}$ variables.
In this study, we use a mutually connected neural network to represent Eq. (\ref{eq13}) using binary variables $\{0,1\}$.
Mutually connected neural networks have been proposed as algorithms for solving various minimization problems. In practice, the traveling salesman problem is solved using this type of neural network in\cite{26}, and can also be applied to the problem defined in Eq. (\ref{eq8}).
The mutually connected neural network and the Ising Hamiltonian have similar energy structures; therefore, after formulating Eq. (\ref{eq8}) in the mutually connected neural network, we can derive the parameters of the Ising Hamiltonian, $J_{ijkl}$ and $\lambda_{ij}$, by converting the output from $\{0,1\}$ into $\{-1,+1\}$.
In the following, we first introduce mutually connected neural networks. We then present how to solve RA in NOMA using a mutually connected neural network. Finally, we present the RA in a NOMA system using the CIM.

\subsection{Mutually Connected Neural Network}
Let us consider a neural network with a two-dimensional structure of $N\times M$; the mutually connected neural network has the following energy structure:

\begin{equation}
\label{eq14}
E(x)=-\frac{1}{2}\sum_{i=1}^{N}\sum_{j=1}^{M}\sum_{k=1}^{N}\sum_{l=1}^{M}{w_{ijkl}x_{ij}x_{kl}}+\sum_{i=1}^{N}\sum_{j=1}^{M}{\theta_{ij}x_{ij}} ,
\end{equation}
where $x_{ij}\in\{0,1\}$ is the state of the $(i,j)$th neuron, $w_{ijkl}$ is the coupling weight between the $(i,j)$th and $(k,l)$th neurons, and $\theta_{ij}$ is the firing threshold of the $(i,j)$th neuron.
From the similarity of the energy structure of the Ising Hamiltonian  and that of the mutually connected neural network, i.e., Eqs. (\ref{eq13}) and (\ref{eq14}), the parameters of the CIM, $J_{ijkl}$ and $\lambda_{ij}$, can be derived using the parameters (i.e., $w_{ijkl}$ and $\theta_{ij}$) of the mutually coupled neural network.
Here, the state of a neuron in a network is defined by the following equation:

\begin{equation}
\label{eq15}
x_{ij}(t+1)=u\left[\sum_{k=1}^{N}\sum_{l=1}^{M}{w_{ijkl}x_{kl}-\theta_{ij}}\right] ,
\end{equation}
where $u$ is the Heaviside step function, namely, $u[y]=0$ for $y\leq0$, and $u[y]=1$ for $y>0$.
The energy function of the mutually connected neural network, i.e., Eq. (\ref{eq14}) always decreases and converges to a locally minimum value by updating the neurons iteratively according to Eq. (\ref{eq15}) if the following conditions are satisfied. 1). The self-connections of all neurons are zero, i.e., $w_{ijij}=0$. 2). The mutual connections are symmetric, i.e., $w_{ijkl}=w_{klij}$.

To solve the optimization problem using a mutually connected neural network, we must formulate the optimization problem as an objective function using a neuron $x_{ij}$.
Then, the formulated objective function is compared with Eq. (\ref{eq14}) to derive $w_{ijkl}$ and $\theta_{ij}$. 
By iteratively updating the neurons according to Eq. (\ref{eq15}) using the derived parameters $w_{ijkl}$ and $\theta_{ij}$, Eq. (\ref{eq14}) converges to a local minimum value, and the states of the neurons at that time correspond to the approximate optimal solution to the optimization problem.
In the next subsection, we define neurons $x_{ij}$ and derive $w_{ijkl}$ and $\theta_{ij}$ for optimizing the RA in a NOMA system.

\subsection{Mutually Connected Neural Network to Solve RA in NOMA System}

To optimize the RA problem in a NOMA system using a mutually connected neural network, we define the binary variables in Eq. (\ref{eq9}) as neurons.
Then, our formulated object function Eq. (\ref{eq8}) can be transformed into the energy function of the mutually connected neural network, which is expressed as follows:

\begin{equation}
  \label{eq19}
\begin{split}
E_1&=\sum_{i=1}^{N_u+N_d}\sum_{j=1}^{N_c}\sum_{{\substack{k=1 \\ k\neq i}}}^{N_u+N_d}{-\left(R_{ik}^j+R_{ki}^j\right)}x_{ij}x_{kj}\\
&=\sum_{i=1}^{N_u+N_d}\sum_{j=1}^{N_c}\sum_{k=1}^{N_u+N_d}\sum_{l=1}^{N_c}{-\delta_{jl}\left(1-\delta_{ik}\right)\left(R_{ik}^j+R_{ki}^l\right)x_{ij}x_{kl}} ,
\end{split}
\end{equation}
where $\delta_{ij}$ is Kronecker's delta, namely, $\delta_{ij}=1$ when $i=j$, and $\delta_{ij}=0$, otherwise.
In addition, the constraints (8.c) and (8.d) can be formulated as constraint terms using neurons, which can be expressed as follows:
\begin{small}
\begin{equation}
  \label{eq20}
\begin{split}
E_2&=\sum_{i=1}^{N_u+N_d}\left(\sum_{j=1}^{N_c}x_{ij}-1\right)^2\\
&=\sum_{i=1}^{N_u+N_d}\sum_{j=1}^{N_c}\sum_{k=1}^{N_u+N_d}\sum_{l=1}^{N_c}{\delta_{ik}\left(1-\delta_{jl}\right)x_{ij}x_{kl}}-\sum_{i=1}^{N_u+N_d}\sum_{j=1}^{N_c}{x_{ij}} ,
\end{split}
\end{equation}
\end{small}
\begin{small}
\begin{equation}
  \label{eq21}
\begin{split}
E_3&=\sum_{j=1}^{N_c}\left(\sum_{i=1}^{N_u+N_d}{x_{ij}-2}\right)^2\\
&=\sum_{i=1}^{N_u+N_d}\sum_{j=1}^{N_c}\sum_{k=1}^{N_u+N_d}\sum_{l=1}^{N_c}{\delta_{jl}\left(1-\delta_{ik}\right)x_{ij}x_{kl}}-3\sum_{i=1}^{N_u+N_d}\sum_{j=1}^{N_c}{x_{ij}} .
\end{split}
\end{equation}
\end{small}
Note that the above constraints term can be fully satisfied even if the number of users in the cell $N_u$ is less than $2N_c$, since we define number of the dummy users $N_d$ to always satisfy $N_u+N_d=2N_c$.
Using $E_1$, $E_2$, and $E_3$, the energy function of the RA optimization problem in a NOMA system can be obtained as follows:
\begin{equation}
  \label{eq22}
E_{NOMA}=\epsilon E_1+\zeta E_2+\eta E_3 ,
\end{equation}
where $\epsilon$, $\zeta$, and $\eta$ are parameters that adjust the scaling of each term, which are used to adjust the objective and constraint terms, causing the energy structure to more likely reach the optimal solution.
By comparing Eq. (\ref{eq22}) with Eq. (\ref{eq14}), the coupling weight $w_{ijkl}$ and firing threshold $\theta_{ij}$ for solving the RA optimization problem in a NOMA system using mutually connected neural networks can be obtained as follows:
\begin{equation}
  \label{eq23}
\begin{split}
w_{ijkl}=&2\{{\epsilon\delta}_{jl}\left(1-\delta_{ik}\right)\left(R_{ijk}+R_{kli}\right)-\zeta\delta_{ik}\left(1-\delta_{jl}\right)\\
&-\eta\delta_{jl}\left(1-\delta_{ik}\right)\} ,
\end{split}
\end{equation}

\begin{equation}
  \label{eq24}
\theta_{ij}=-\left(\zeta+3\eta\right).
\end{equation}
Using $w_{ijkl}$ and $\theta_{ij}$ obtained in Eqs. (\ref{eq23}) and (\ref{eq24}), the neurons were updated according to Eq. (\ref{eq15}). Thereby, the combination of neurons minimizes the energy function in Eq. (\ref{eq14}), i.e., the optimal solution to the NOMA RA problem can be obtained. However, owing to the monotonically decreasing nature of the energy function, this neural network is likely to fall into a local minimum of the network state. Hence, the optimal solutions may not be achieved using the mutually connected neural network. Further, it has been shown that the Ising Hamiltonian near the optimal solution can be reached with high probability by amplifying the DOPO pulse to near the oscillation threshold in CIM \cite{21b}. Therefore, in the next subsection, we will use the derived parameters $w_{ijkl}$ and $\theta_{ij}$ to derive $J_{ijkl}$ and $\lambda_{ij}$ which are the parameters used to solve the optimization problem in the CIM.

\subsection{CIM Used to Solve RA in NOMA System}
As previously noted, the Ising Hamiltonian of the CIM and the energy function of the mutually connected neural network have similar energy structures.
As a difference, the neuron $x_{ij}\in \{0,1\}$ in the mutually connected neural network has states of 0 and 1, whereas the Ising-spin $\sigma_{ij}\in \{-1,+1\}$ in the Ising model has states of -1 and +1. 
To solve the RA problem using the CIM, we redefine the output of the neurons in a mutually connected neural network as -1 and +1.
Using the neuron $\tilde{x}_{ij}\in \{-1,+1\}$ with a redefined output, the update equation for the neuron is expressed as follows:

\begin{equation}
  \label{eq25}
\tilde{x}_{ij}(t+1)=r\left[\sum_{k=1}^{N}\sum_{l=1}^{N}{\tilde{w}_{ijkl}\tilde{x}_{kl}-\tilde{\theta}_{ij}}\right] ,
\end{equation}
where $r[y]=-1$ for $y\leq0$, and $r[y]=+1$ for $y>0$, and $\tilde{w}_{ijkl}$ and $\tilde{\theta}_{ij}$ are the coupling weight and firing threshold of the neural network with output $\{-1,+1\}$, respectively.
We then can obtain the following equation by transforming the internal state of Eq. (\ref{eq15}) using $\tilde{x}_{ij}=2x_{ij}-1$:
\begin{equation}
  \label{eq26}
\begin{split}
        &\sum_{k=1}^{N}\sum_{l=1}^{N}{w_{ijkl}x_{kl}-\theta_{ij}}\\
        &=\sum_{k=1}^{N}\sum_{l=1}^{N}{\frac{w_{ijkl}}{2}\tilde{x}_{kl}}-\left(\theta_{ij}-
        \sum_{k=1}^{N}\sum_{l=1}^{N}{\frac{w_{ijkl}}{2}}\right) .
\end{split}
\end{equation}
By comparing Eqs. (\ref{eq14}) and (\ref{eq26}), we can obtain the following equations for $\tilde{w}_{ijkl}$ and $\tilde{\theta}_{ij}$, respectively.
\begin{equation}
  \label{eq27}
    \tilde{w}_{ijkl}=J_{ijkl}=\frac{w_{ijkl}}{2} ,
\end{equation}
\begin{equation}
  \label{eq28}
    \tilde{\theta}_{ij}=\lambda_{ij}=\theta_{ij}-\sum_{k=1}^{N}\sum_{l=1}^{N}{\frac{w_{ijkl}}{2}} .
\end{equation}
By setting these derived parameters in the FPGA module, a fast RA optimization of the NOMA system using the CIM can be performed.

\begin{figure*}[!t]
    \centering
    \includegraphics[width=170mm, clip]{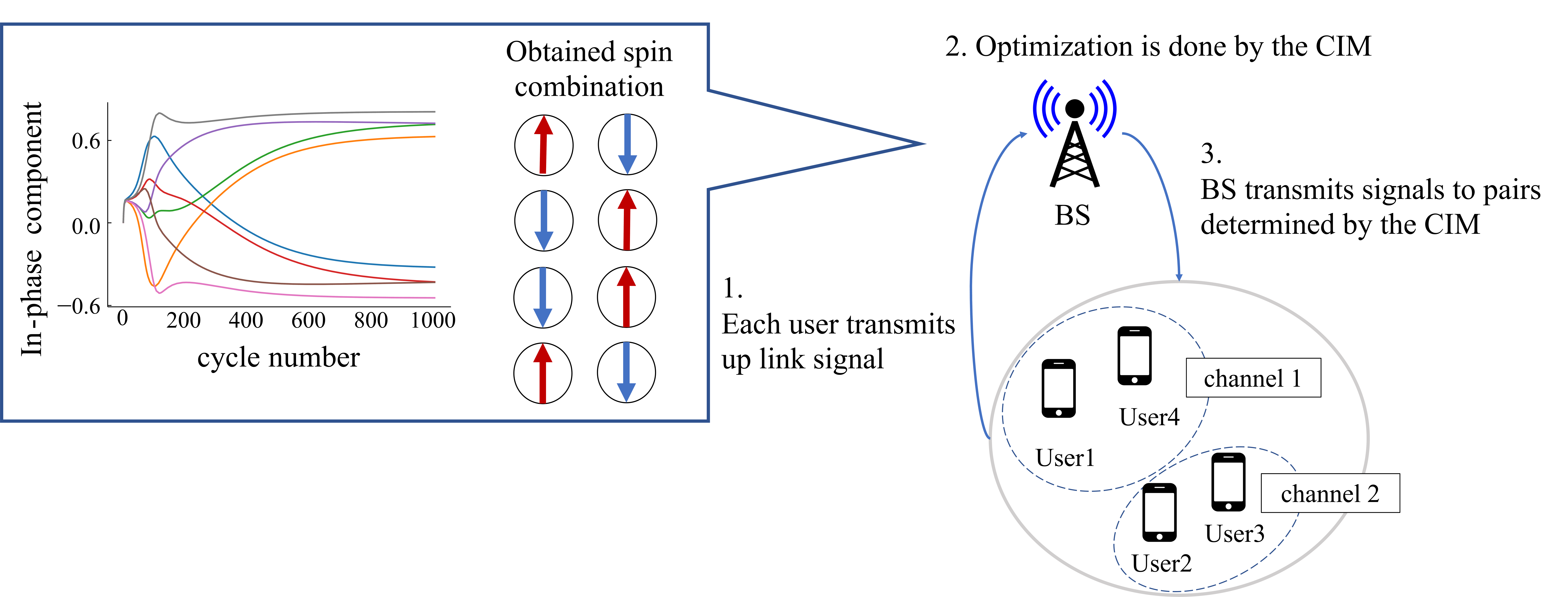}
    \caption{Overview of our proposed CIM-based RA method
    }
    \label{fig:operation}
\end{figure*}

Then, we present the details of the optimization procedure of RA using the CIM in NOMA systems.
In this study, the BS performs channel allocation using the CIM for the NOMA system to achieve the optimal total reachable data rate.
A schematic of the optimization procedure of the NOMA system is shown in Fig. \ref{fig:operation}.
For simplicity, we present a NOMA system with 4 users and 2 channels as an example.
As shown in Fig. \ref{fig:operation}, each user transmits the data to the BS. Then, the channel assignment is determined by the CIM equipped in the BS.
There are DOPO pulses on the optical fiber of the CIM, each of which behaves as an Ising spin.
Because this problem involves 4 users and 2 channels,
the RA problem can be solved using the CIM implemented with $2\times4=8$ DOPO pulses on the fiber.
When the CIM is operated, the states of the in-phase components shown on the left side of Fig. \ref{fig:operation}  are obtained.
Here, the cycle number represents the number of times the DOPO pulse travels around the optical fiber of the CIM.
As the cycle count increases, the in-phase component of the amplitude of the DOPO pulses increases and eventually splits into positive or negative values, i.e., 0 phase or $\pi$ phase.
In the CIM, the spin orientation is determined by the in-phase components of the amplitude.
Here, when user $i$ uses channel $j$, the spin is +1, i.e., an upward spin.
Thus, the obtained spin combination corresponds to the channel assignment.
Finally, based on this channel assignment, the BS sends a signal to the users after assigning the optimal power using Eq. (\ref{eq11}).
The pseudo-code for the algorithm is presented as Algorithm \ref{alg1}.

\begin{algorithm}[t]
    \caption{Optimization of RA Using the CIM in NOMA System}
    \label{alg1}
    \begin{algorithmic}[1]
        \REQUIRE {Distance between user $i$ and the BS $d_i$, and the channel coefficient of user $i$ in channel $j$ $g_i^j$}
        \ENSURE {channel allocation $\sigma_{ij}$, power allocation $q_j$, $P_{ik}^j$.}
        \STATE \textbf{Initialize:}
        \STATE \hspace*{\algorithmicindent}\parbox[t]{0.9\linewidth}{\raggedright{
                       Initialize the channel response $h_i^j$ and the CNR $\Gamma_i^j$ according to (\ref{eq3_1}), (\ref{eq3_2})
        }}
        \STATE \hspace*{\algorithmicindent}\parbox[t]{0.9\linewidth}{\raggedright{
            Set the power allocation $P_{ik}^j$ according to (\ref{eq11}) with $q_j=1.0\ \ \forall j$.
        }}
        \STATE \hspace*{\algorithmicindent}\parbox[t]{0.9\linewidth}{\raggedright{
            Calculate data rate for users according to (\ref{eq6a}).
        }}
        \STATE \textbf{Channel allocation by the CIM:}
        \STATE \hspace*{\algorithmicindent}\parbox[t]{0.9\linewidth}{\raggedright{
            Calculate the $J_{ijkl}$ and $\lambda_{ij}$ according to (\ref{eq27}), (\ref{eq28}) and setting these parameters to FPGA.
        }}
        \STATE \hspace*{\algorithmicindent}\parbox[t]{0.9\linewidth}{\raggedright{
            Obtaining spins that minimizes the Ising Hamiltonian $\sigma_{ij}$ by the CIM as described in Section IV.
        }}  
        \STATE \textbf{Optimal power allocation:}    
        \STATE \hspace*{\algorithmicindent}\parbox[t]{0.9\linewidth}{\raggedright{
             $q_j$ is obtained by water filling algorithms \cite{21a} and
            re-calculate $P_{ik}^j$.
        }}
        
    \end{algorithmic}
\end{algorithm}

\subsection{Complexity Analysis of the CIM}
In this subsection, we discuss the time and computational complexity of the proposed method. In [26], it is demonstrated that the CIM takes a constant amount of time to obtain the ground state as long as the fiber length is constant. That is, the time complexity of the CIM is $\mathcal{O}(1)$ experimentally. The reason is that the computation time of one cycle for the CIM is related to the length of a fiber. For the CIM with 2000 spins and 1km fiber, 1000 cycles running of the CIM can be achieved within 5 ms while the spins can converge to the ground state within 1000 cycles by amplifying pump pulse amplitude appropriately. In our paper, the problems can be solved using the CIM with 2000 spins. Hence, the computation time is within 5 ms.

Then, we discuss the computation complexity of the main computation part of the CIM, i.e., FPGA. Ref. [16] shows that the memory resources required for matrix computation and power consumption in FPGAs are scaled by $\mathcal{O}(N^2)$. A typical CPU runs 10-100Gflop per second, so its power efficiency is often below 1 Gop/J. Similarly, GPUs also suffer from high power consumption. Hence, CPUs and GPUs are difficult to apply to applications that require low power consumption \cite{39}. In addition, although parallel processing is possible for modern CPUs with multi-core, unnecessary parts also need to be parallelized since the entire core is used. In the case of FPGA, only necessary processing needs to be programmed. Hence, parallel processing can be performed without wasting computation resources, increasing the computation efficiency. Moreover, owing to the merit of the parallel processing of the FPGA, the FPGA uses less circuit area when performing the same processing compared to the CPU/GPU and can perform processing without resource waste. It is natural with high power efficiency. In summary, our proposed method may be superior to the resource allocation method computed using the traditional computer in computation complexity.

\section{Performance Evaluation}
\label{sect:simulationresult}

In this section, we evaluate the performance in the convergence, data rate, and computation time of our proposed CIM-based RA for a NOMA system and compare it with the SA (a meta-heuristic optimization method), a conventional-NOMA pairing scheme (C-NOMA)\cite{5}, a DRL and the ES methods.
Specifically, the RA is applied as follows. First, the channels are allocated using each method.
The optimal power allocation is then applied using Eqs. (\ref{eq11}) and (\ref{eq12}). The following of this section is organized as follows. First, the simulation settings is introduced. Then, the comparison methods are presented. Next, the performance in convergence, data rate, computation time of our proposed method and the comparison with the other methods are evaluated.

\subsection{Simulation Settings}
In the simulation, a circular cell of a NOMA system with a radius of 500 m is considered. A BS is placed at the center of the cell. Users are randomly placed in the cell.
Table II shows the detailed parameter setting used in our simulation.
\begin{table}[hbtp]
\centering
\caption{Simulation parameters}
\scalebox{1.2}{
\begin{tabular}{c|c}
Parameters                             & Values      \\ \hline\hline
Total Bandwidth $B$                    & 5.0 MHz     \\
Minimal data rate of each users $R_{min}$ & 2 bps/Hz    \\
Cell radius                            & 500 m        \\
Minimal distance between user and BS       & 50 m        \\
Maximum number of multiplexes users    & 2 users     \\
Noise power spectral density           & -170 dBm/Hz
\end{tabular}
}
\end{table}

\begin{figure*}[!t]

\centering
	\begin{minipage}{.30\textwidth}
		\centering
		\includegraphics[width=5.5cm]{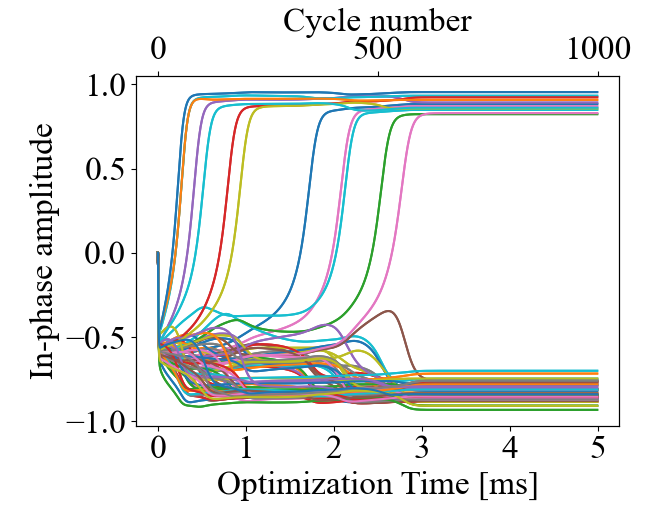}
	\end{minipage}
	\begin{minipage}{.32\textwidth}
		\centering
		\includegraphics[width=6cm]{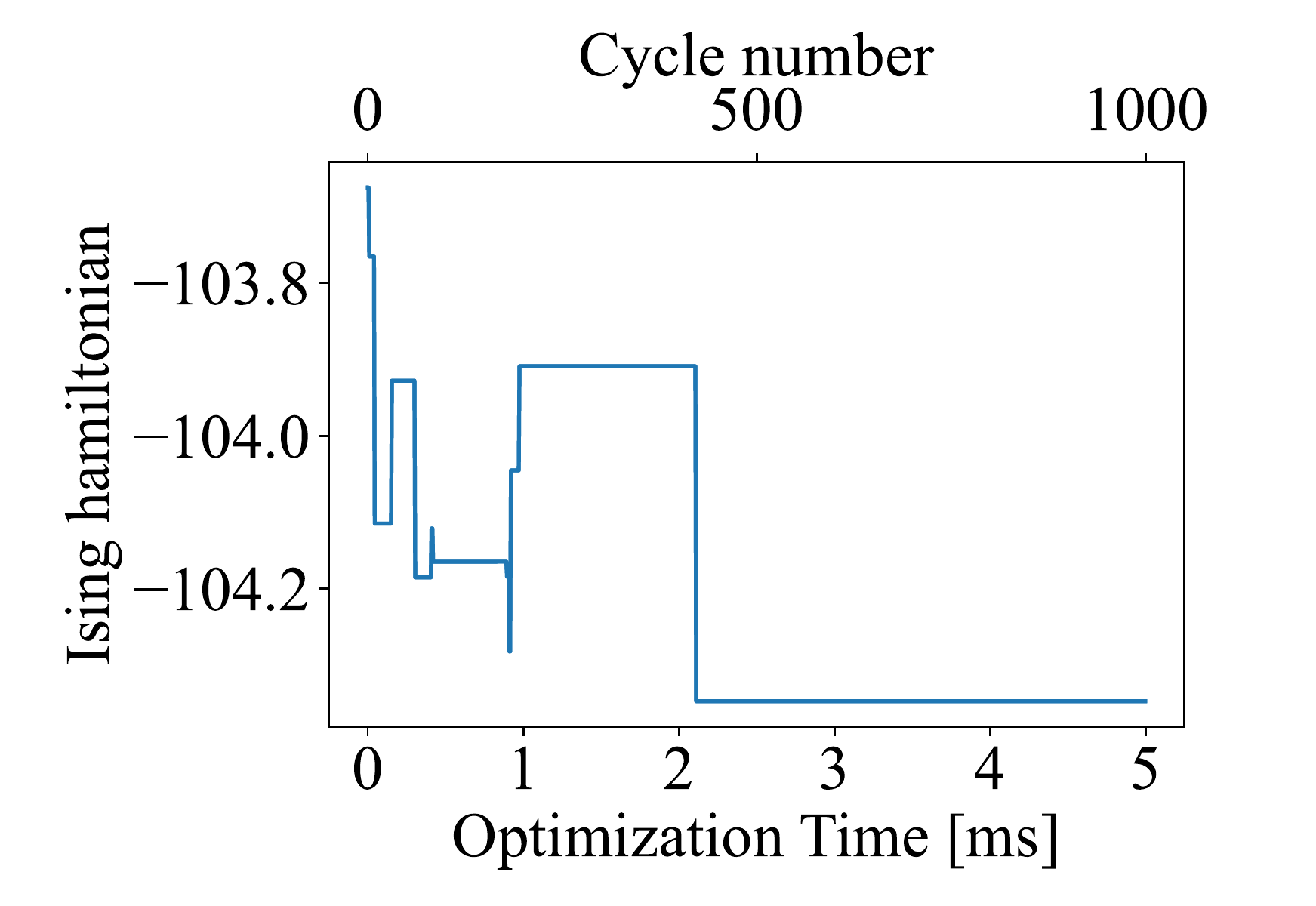}
	\end{minipage}
	\begin{minipage}{.30\textwidth}
		\centering
		\includegraphics[width=5.5cm]{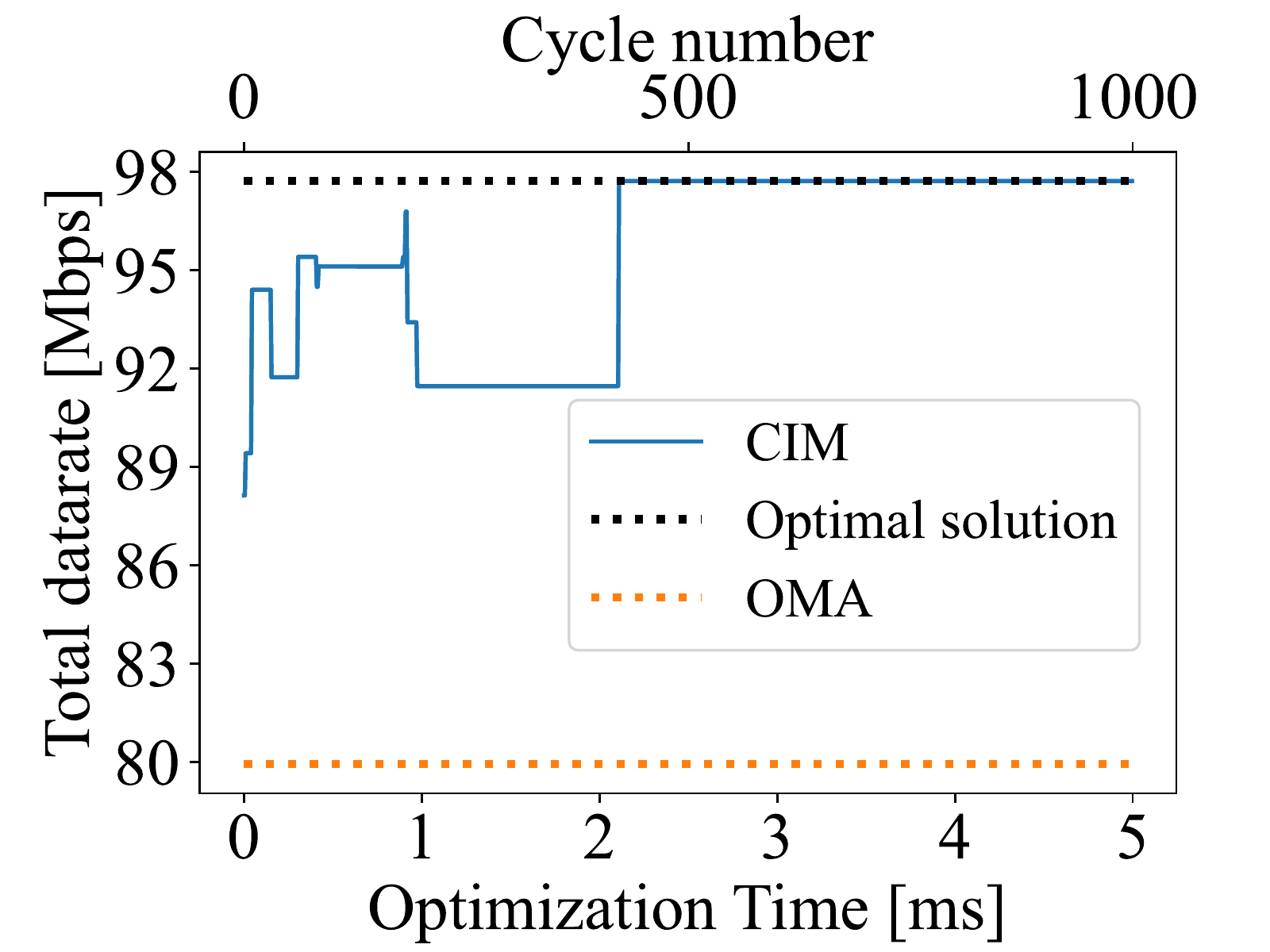}
	\end{minipage}
	
\caption{Convergence of in-phase components of DOPO pulse (left) and the corresponding Ising Hamiltonian at that time (center) and comparison of the total data rate obtained by the CIM-based RA with the optimal and the OMA methods at that time (right) in the NOMA system with $N_u=12$ users and $N_c=6$ sub-channels.}
	
	\label{spin_convergence}
\end{figure*}

We use the following simulation model of the CIM to evaluate our proposed method:
 \begin{equation}
 \label{eq6_1}
\frac{dc_{ij}}{dt}=\left(-1+p-c_{ij}^2-s_{ij}^2\right)c_{ij}+\sum_{k=1}^{N}\sum_{l=1}^{N}{J_{ijkl}c_{kl}}-\lambda_{ij} ,
\end{equation}
\begin{equation}
\label{eq6_2}
\frac{ds_{ij}}{dt}=\left(-1-p-c_{ij}^2-s_{ij}^2\right)s_{ij}+\sum_{k=1}^{N}\sum_{l=1}^{N}{J_{ijkl}s_{kl}}-\lambda_{ij} ,
\end{equation}
where $c_{ij}$ and $s_{ij}$ are the in-phase and quadrature-phase amplitude of the $(i,j)$th optical pulses, respectively, and $p$ is the pump pulse, which is used to amplify $c_{ij}$.
By running the simulation with a gradual increase in $p$, the in-phase and quadrature components of the DOPO pulse can be obtained.
As described in Section \Rmnum{4}, the measurement feedback of the CIM reproduces the Ising spins using the amplitudes of DOPO pulses cycling over long fibers.
In other words, the simulation model can be used to reproduce the behavior of the CIM and obtain the combination of spins that minimizes the Hamiltonian.
Ising spin $\sigma_{ij}$ is implemented by $c_{ij}$: If $c_{ij}<0$, then $\sigma_{ij}=-1$, 
whereas if $c_{ij}>0$, then $\sigma_{ij}=+1$.

\subsection{Compared Methods}

\subsubsection{SA-based RA}
In SA-based RA, the Boltzmann machine model\cite{27} is used as a computational model. The Boltzmann machine is a type of mutually connected neural network that incorporates the statistical behavior. Specifically, the output function of the mutually connected neural network is a sigmoid function with temperature $T$ introduced, 
and the probability that the value of neuron $X\in\{0,1\}$ changes with the temperature $T$. In our simulation, the initial temperature of the SA was set as $T_{ini}=5.0$, and is cooled according to the following equation for each iteration $t$ as described in \cite{28}:
\begin{equation}
T(t)=\frac{T_{ini}}{log(1+t)} ,
\end{equation}
where $T(t)$ is the temperature of the SA at iteration $t$.
The number of SA iterations is measured as 10,000.

\subsubsection{C-NOMA}
In the C-NOMA, users are divided into a set of near-users that are located near the center of the cell and far-users that are located far from the center of the cell based on criteria such as the distance or channel gain.
Thereafter, channels are allocated by pairing one user from the near-user set and one user from the far-user set, and assigning them to the same channel.
\subsubsection{DRL-based RA method}
As a DRL-based RA method for comparison, we designed a DQN method to solve our formulated RA optimization problem shown in Eq. (\ref{eq8}). In the designed DQN method, 
channel assignment is based on the following information at the decision time $t$. CSI matrix information  $\textbf{CSI}_t \in \mathbb{R}^{N_u\times N_c}$, user assignment variables $\textbf{x}_t \in \mathbb{R}^{N_u\times N_c}$,
user to be assigned to the channel in the next step $\textbf{A}_t \in \mathbb{R}^{N_u}$,
the achieved data rate of each user in the last step $\textbf{Rate}_t \in \mathbb{R}^{N_u}$.
That is, the state at the decision time $t$, which is denoted as $S_t$ is defined as
$S_t=[\textbf{CSI}_t, \textbf{x}_{t-1}, \textbf{A}_t, \textbf{R}_{t-1}]$.
Here, the action is defined as the channel allocation, i.e., the channel allocated to which user.
The policy used in our designed method is $\epsilon$-greedy method while the reward is defined as the data rate \cite{21}.
The Q network and Target network of DQN consist of 3 layers, and the size of each layer is $|S_t|$, 256, and $N_c$, respectively.
RMSProp is used to update the parameters of neural network, and the learning rate is set to $10^{-4}$.
\subsubsection{ES} In the ES, the channel allocation is the optimal solution by exhaustive searching.

\begin{figure*}[t]
\centering
	\begin{minipage}{.49\textwidth}
		\centering
		\includegraphics[width=9cm]{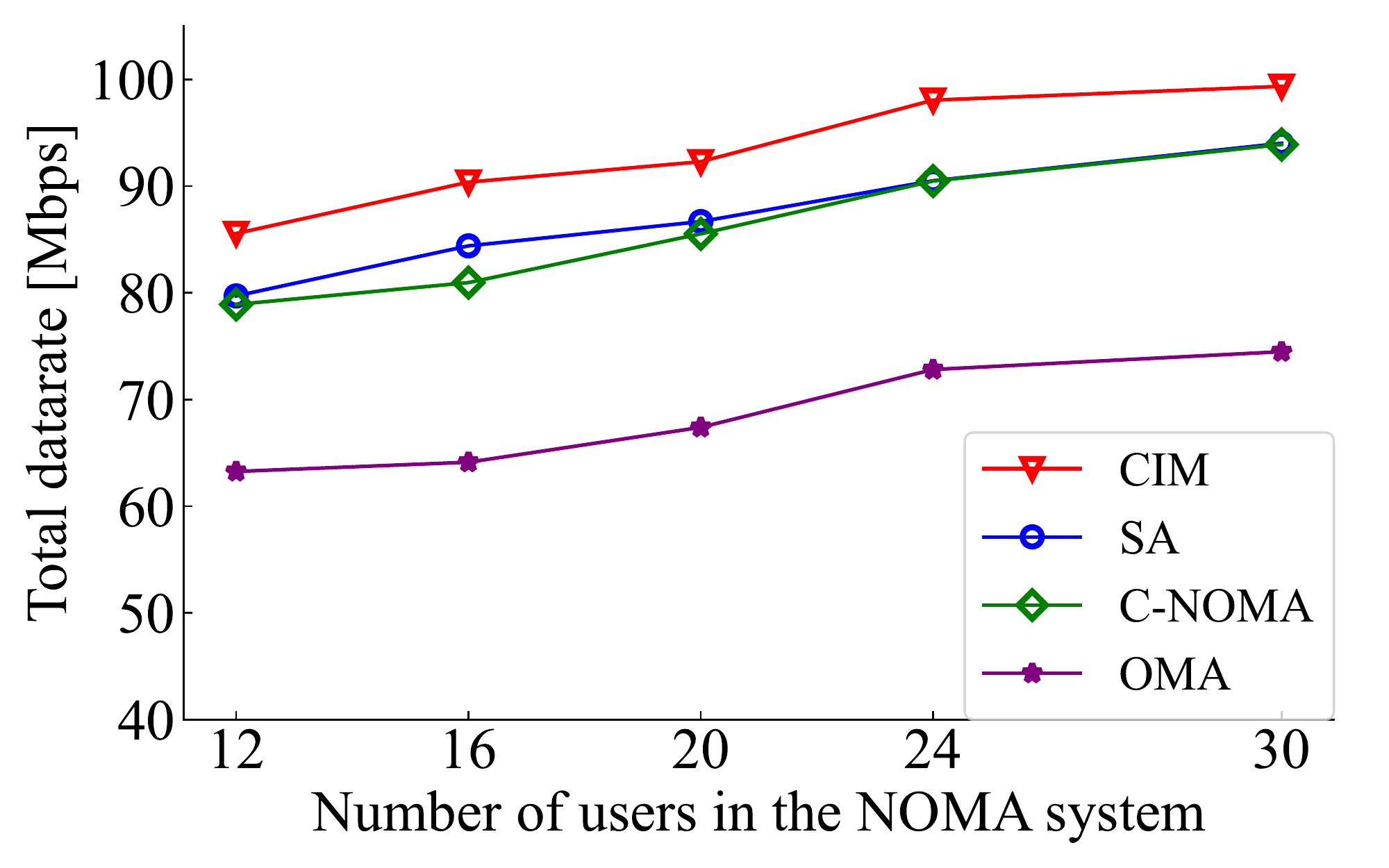}
		\caption{Total data rate versus number of users $N_u$ in the NOMA system with total transmit power $P_T=12$ from the BS.}
		\label{fig:result1_user and channel increases}
	\end{minipage}
		\begin{minipage}{.49\textwidth}
		\centering
		\includegraphics[width=9cm]{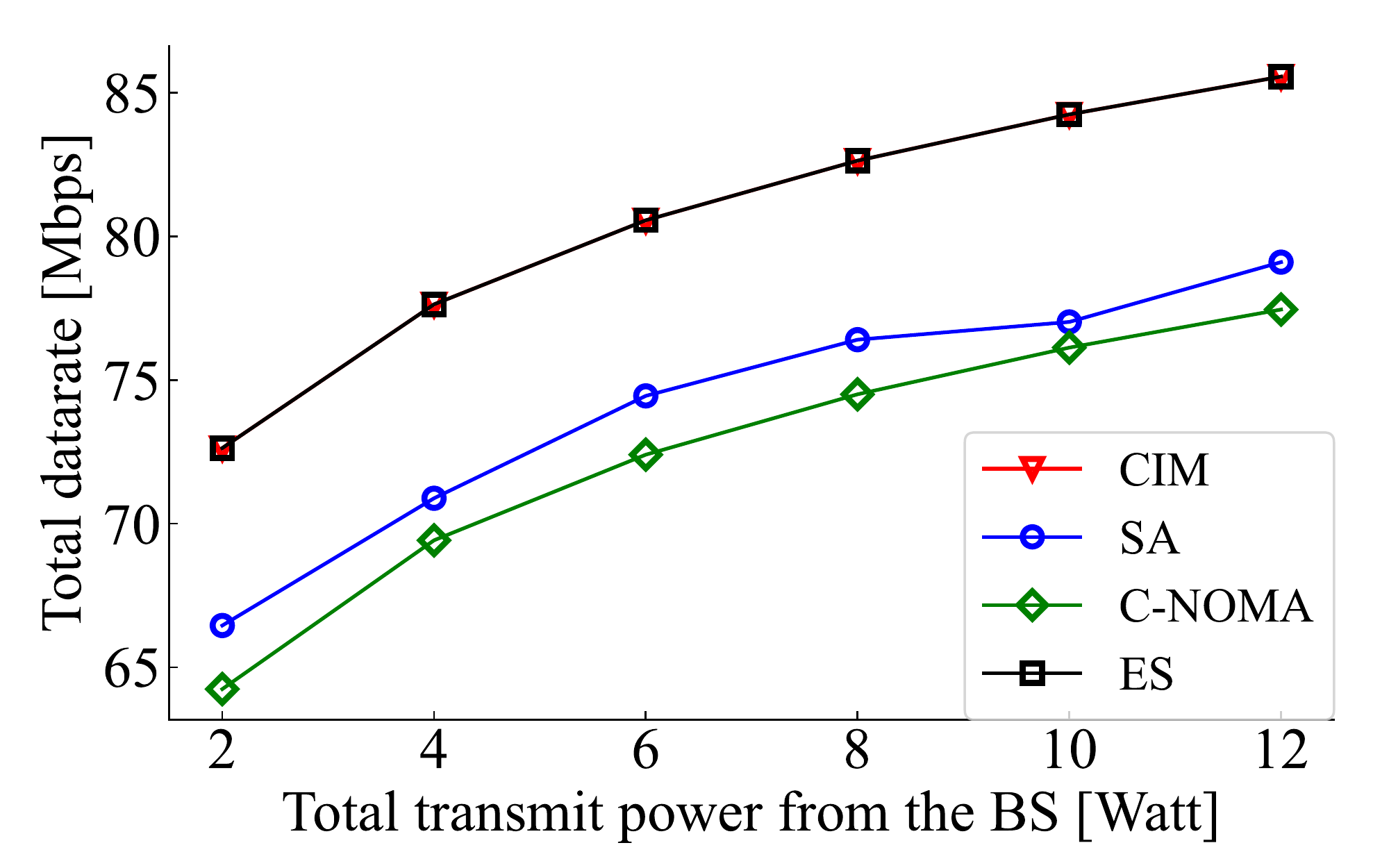}
		\caption{Total data rate versus total transmit power $P_T$ from the BS with number of users $N_u=12$ and number of sub-channels $N_c=6$ in the NOMA system.}
		\label{fig:result2_watt increases}
	\end{minipage}
\end{figure*}

\subsection{Convergence of the Proposed Method}
First, we evaluated the convergence of the proposed method in the NOMA system with 12 users and 6 subchannels, i.e., $N_u=12$ and $N_c=6$.
Figure \ref{spin_convergence} shows the time evolution of the in-phase components of the DOPO pulses for each cycle and the corresponding energy of the Ising Hamiltonian at that time.
Here, for the optimization time, we used the time required for 1000 cycles on an actual CIM machine as shown in \cite{10}.

The subfigures on the left side, in the center, on the right side of Fig. \ref{spin_convergence} are the in-phase amplitude of the spins, the corresponding Ising Hamiltonian, and the total data rate of the three RA methods in the NOMA system, respectively. From Fig. \ref{spin_convergence}, we can see that when the in-phase amplitude of the spins converges, the minimum Ising Hamiltonian reaches, while the total data rate of the CIM-based RA achieves the optimal value. This means that in the process of the CIM operations, the spins spontaneously selected a combination of spins that exhibited the ground state of the Ising Hamiltonian, which corresponds to the solutions to the formulated optimization problem. Hence, the convergence of our proposed method has been verified.

In addition, it is shown in \cite{12a} that for certain MAX-CUT problems, the real CIM machines can provide a stable state approximately 1000 times faster than SA using state-of-the-art CPUs.
Thus, our proposed method has a significant advantage in convergence speed.
Moreover, in \cite{37}, it has been demonstrated that the phase of the DOPO pulse during oscillation converges to the ground state within 5 ms by amplifying the DOPO pulse with an appropriate pump pulse amplitude. The setting of the pump pulse amplitude in real CIM is out of the scope of this manuscript, while considering the space of the manuscript, we omit the detailed analytical proof process, which can be found in [19], [32], \cite{37}. In this paper, we set the pump pulse amplitude, i.e., $p$ in equations (\ref{eq6_1}) and (\ref{eq6_2}), to an appropriate value by parameter search during the simulations. Thus, the in-phase amplitude of the DOPO pulse can converge within 5ms, as shown in Fig. 4.

\subsection{Performance Evaluation of the Data Rate under Static Environment}
\subsubsection{Data rate vs the number of users}
Then, we evaluated the total data rate when the number of users $N_u$ and that of subchannels $N_c$ are varied. Note that we consider that up to 2 users are multiplexed in each subchannel. Therefore, as $N_c$ increases, $N_c$ also increases as $N_c=\lceil N_u/2 \rceil$. In other words, $N_c$ is varied from 12 to 30 in this simulation, while at the same time $N_c$ is varied from 6 to 15.
In addition, the transmission power $P_T$ is set to 12. 
Figure \ref{fig:result1_user and channel increases} shows the simulation results. From Fig. \ref{fig:result1_user and channel increases}, we can see that our proposed method obtained the highest total data rate among the compared methods.
This indicates that the most suitable joint channel and power allocation is possible using the proposed method.
In addition, we can observe that the NOMA methods achieve a higher total data rate than the existing OMA method.
During this simulation, we set up to 30 users in the NOMA system.
As $N_u$s increased, the number of combinations of user-channel pair increased, making it difficult to obtain an optimal solution using the ES.
By contrast, $N_c \times N_u=450$ spins are required when solving this RA problem with 30 users and 15 channels using the CIM.
Because the actual CIM can currently solve problems with up to 100,000 spins\cite{12a}, this RA problem can be solved using the CIM.
In other words, using the CIM, it is possible to obtain an effective solution to the RA optimization problem in a large-scale NOMA system that cannot be solved using the ES at a speed on the order of milliseconds.

\subsubsection{Data rate vs transmit power}Next, we evaluated the total data rate of the NOMA system when varying the total transmission power from the BS, i.e., $P_T$. In this simulation, the numbers of users and sub-channels in the NOMA system are set to $N_u=12$ and $N_c=6$, respectively. Here, $P_T$ is set to 2, 4, 6, 8, 10, and 12 Watts.
Figure \ref{fig:result2_watt increases} shows the simulation results.
From Fig. \ref{fig:result2_watt increases}, we can see that our proposed CIM-based method can obtain a reachable optimal solution regardless of the transmission power from the BS $P_T$.



\begin{figure*}[!t]
\centering
	\subfigure[$\alpha=3.0$]{\includegraphics[width=8cm]{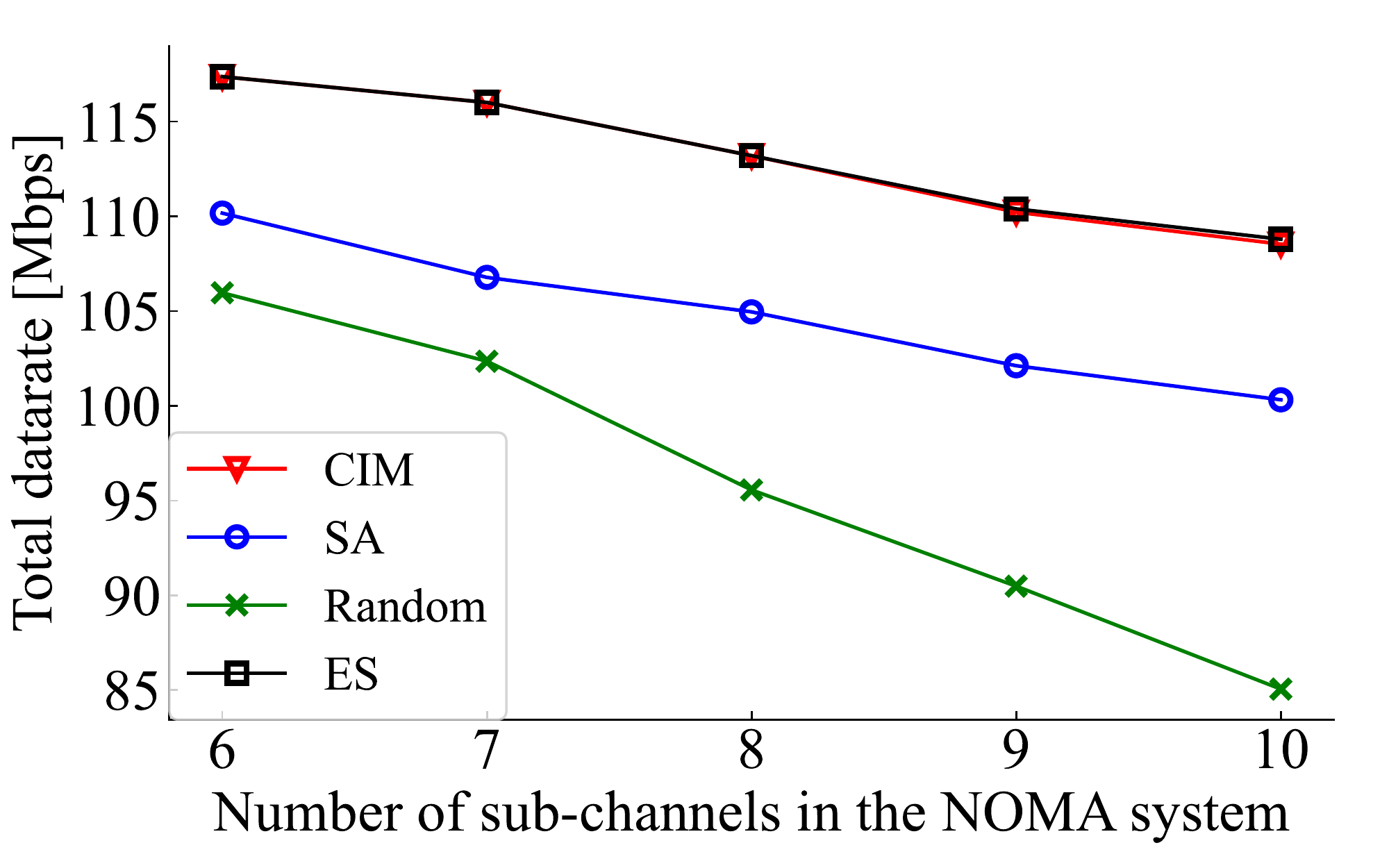}}
        \subfigure[$\alpha=4.0$]{\includegraphics[width=8cm]{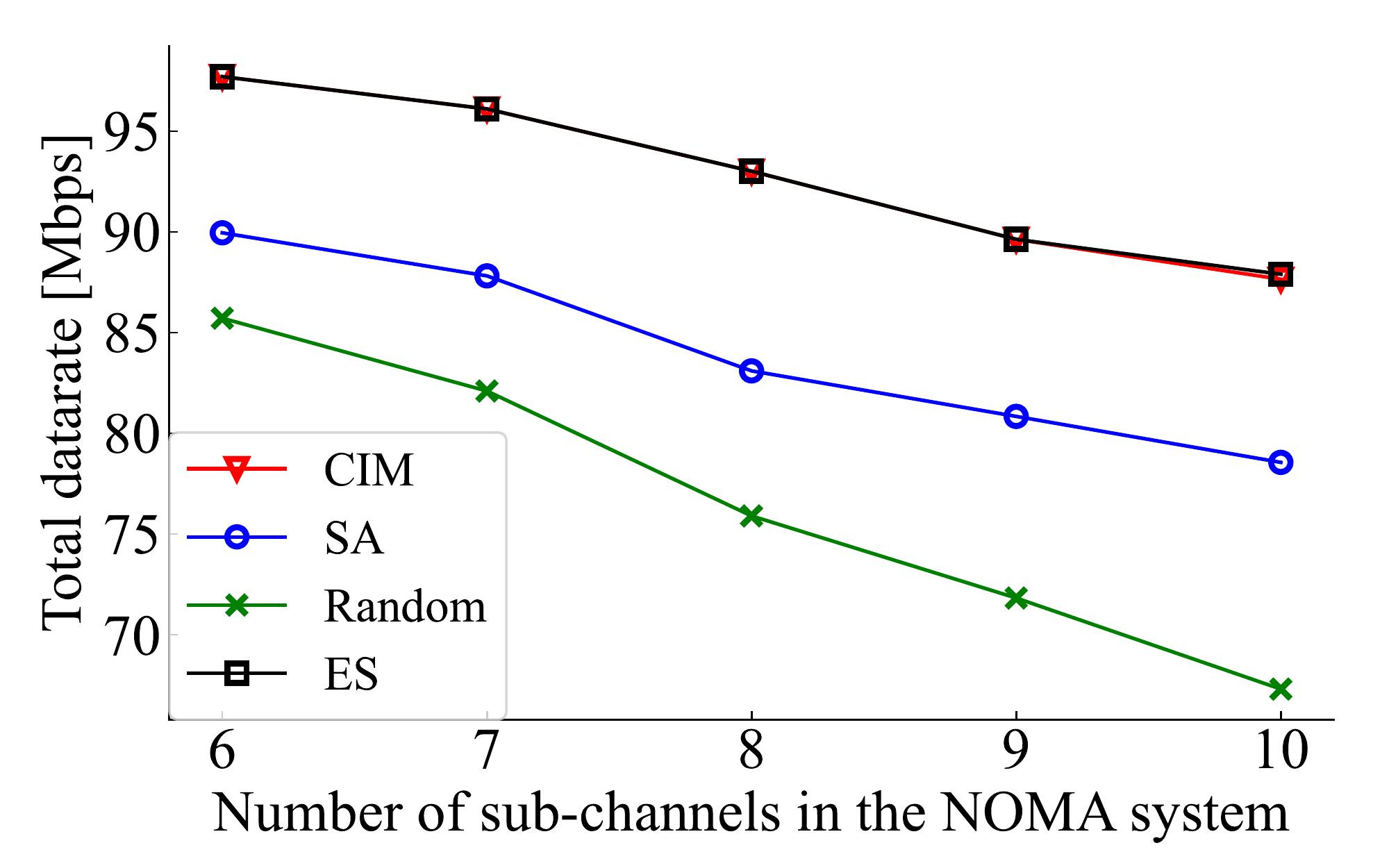}}
\caption{Total data rate versus number of sub-channels $N_c$ with number of users $N_u=12$ in the NOMA system. \\Note that the sub-carrier width changes as the number of sub-channels changes.}
	\label{fig:result3_subchannels increase}
\end{figure*}

\begin{figure*}[!t]
\centering
	\subfigure[$\alpha=3.0$]{\includegraphics[width=8cm]{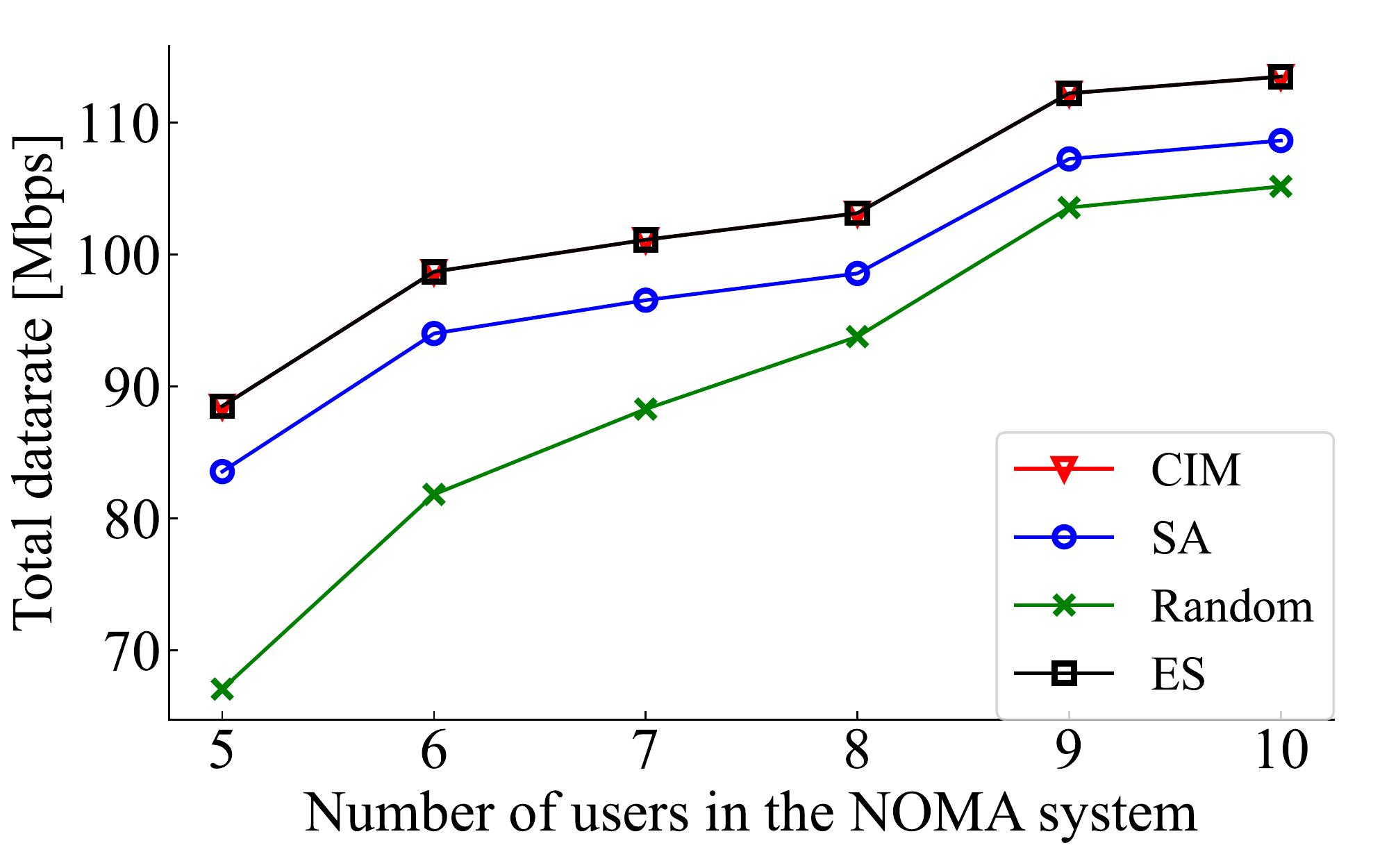}}
        \subfigure[$\alpha=4.0$]{\includegraphics[width=8cm]{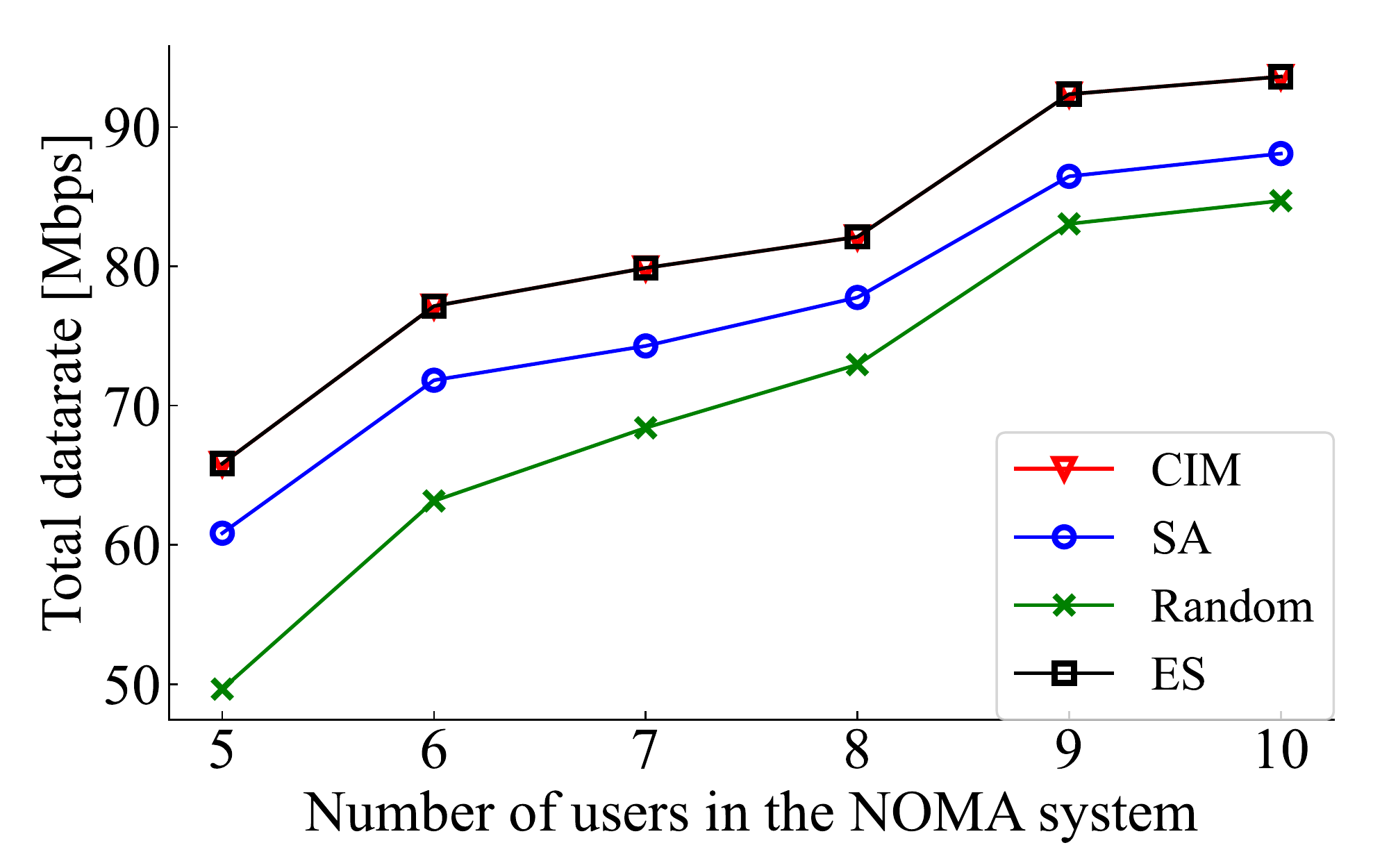}}
\caption{Total data rate versus number of users $N_u$ with number of sub-channels $N_c=5$ in the NOMA system. }
	\label{fig:result4_number of users increase}
\end{figure*}

\subsubsection{Data rate vs the number of channels}Next, we evaluated the total data rate for varying number of channel $N_c$. In this simulation, the transmission power from the BS is set to $P_T=12$. The number of users is set to $N_u=12$. $N_c$ is varied from 6 to 10.
In this study, the total bandwidth $B=5.0$ MHz is constant, and thus the sub-carrier width is $5.0/Nc$ MHz. Note that this means that the sub-carrier width changes as the number of sub-channels changes.
Figure \ref{fig:result3_subchannels increase} shows the simulation results when the path loss coefficient $\alpha$ is set to 3 and 4. From Fig. \ref{fig:result3_subchannels increase} , we can see that the proposed method can achieve the same total data rate as the ES even when $N_c$ increases. 
Hence, we can say that the performance of the proposed method is comparable to that of the ES.
However, when $N_c$ increases, the number of channel-user combinations highly increases. At that time, the total data rate of the random allocation method decreases rapidly.
This indicates that the optimization using the CIM can result in an effective channel allocation.
In addition, the total data rate decreases as $N_c$ increases.
One of the reasons for this is that the power allocated per channel decreases as $N_c$ increases because the power from the BS is constant.
Another reason to be considered can be summarized as follows. When $N_c$ increases, the bandwidth for each sub-channel become small, whereas the OMA is applied in more channels. By contrast, when $N_c$ decreases, the bandwidth for each sub-channel increases while NOMA is applied in more channels. Because the NOMA technique is more efficient than the OMA technique, the total data rate decreases with the increase in the number of sub-channels.

\subsubsection{Data rate vs the number of users under fixed number of channels}Then, we evaluated the total data rate for the varying number of users $N_u$. In the simulation,
$N_u$ is varied from 5 to 10. The number of sub-channels is also set to $N_c=5$. Under this setting, the OMA and NOMA coexist when $N_u$ is less than 10.
The transmission power from the BS is then set to $P_T=12$.
figure \ref{fig:result4_number of users increase} shows the simulation results when the path loss coefficient $\alpha$ is set to 3 and 4. 
From Fig. \ref{fig:result4_number of users increase}, we can see that the proposed method can achieve the same total data rate as the ES even when $N_u$ increases. This indicates that the CIM can achieve the optimal channel selection even when the NOMA and OMA coexist.
In addition, the total data rate of the system increases as $N_u$ increases.
As the reason for this, as $N_u$ increases, the number of channels to be multiplexed increases.
In other words, the advantages of the NOMA system were fully utilized.

\begin{figure}[!t]
    \centering
    \includegraphics[width=90mm, clip]{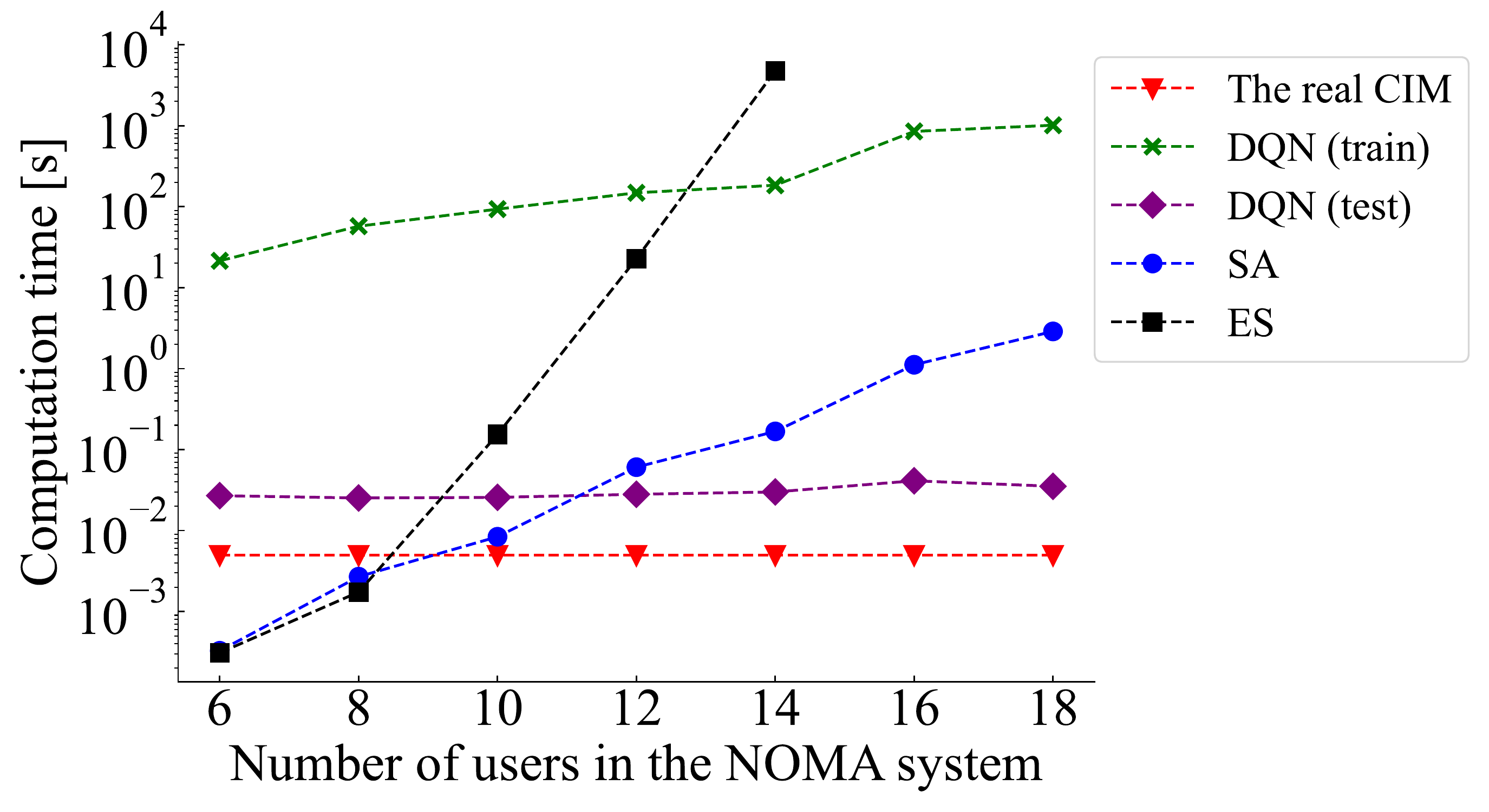}
    \caption{Comparison of computation time.}
    \label{fig:computation_time}
\end{figure}

\subsection{Comparison of the Computation Time}
Then, we evaluate the computation time of the SA, the ES, the DQN-based, and the CIM-based RA method under the varying number of users. Here, the computation time of the SA and the ES is evaluated using an Intel Xeon Gold 5222 CPU @ 3.80 GHz and C language, and the DQN-based RA is evaluated using the same CPU and Python with PyTorch, respectively. The number of users in the NOMA system $N_u$ is set as 6, 8, 10, 12, 14, 16, and 18 in the performance evaluation. The numbers of SA iterations under corresponding different $N_u$ were set as 10, 50, 100, 500, 1000, 5000, and 10000 times, respectively. In addition, the number of training episodes for the DQN-based RA under different $N_u$ is set as 100, 250, 400, 550, 700, 850, and 1000 times, respectively. Here, each episode consists of 120 steps. The BS (Base Station) collects information and trains the neural network for each step. The information consists of the CSI information, the channel assignment information, and the achieved data rate of the previous step of each user. The reason for the settings of the SA and DQN-based RA is that the larger the scale of the problem is, the larger number of iterations and training episodes required for SA and DQN-based RA methods. Here, owing to the enormous amount of computation time involved for the ES with the growth of the number of users, the method was evaluated for up to 14 users. Fig. 9 shows the evaluation results. The computation time of the real CIM in Fig. 9 is the time cost for RA using the CIM in the real world. In [16], it has been shown that the Ising problem can be solved using the CIM in 5 ms regardless of the problem size when the length of the fiber is 1-km-long. The computation time of DQN (train) and DQN (test) represents that for training the neural network and testing (making decisions) using the trained neural network, respectively. As shown in Fig. 9, the CIM-based and DQN-based RA for testing methods demonstrate almost constant computation time even as the number of users in the system increases. In contrast, other algorithms require a much longer time as the problem size increases. However, the DQN-based RA method requires training to obtain a good solution when the communication environment changes. The training time increases with the number of users in the NOMA system. Thus, as described above, the CIM is superior in speed compared to other methods.

\subsection{Performance Evaluation of the Data Rate under Dynamic Environment}
Finally, we evaluate the performance of the proposed method when more realistic user activation is considered. In the simulation, the user activity follows the setup for the urban case in 3GPP TR 36.885 \cite{v2v_useractivity,v2v_useractivity2}. In this scenario, the initial locations of the users are randomly generated according to the spatial Poisson process. Each user drives on the road at a constant speed. The speed of each user is generated randomly between 36km/h to 54km/h. The number of lanes is 2 in each direction and 4 in total in each street. The road grid size by the distance between intersections is 433m*250m. The simulation area size is 1299m*750m. When the user moves to the intersection, the direction is changed with a certain probability that follows the uniform distribution. 
In the performance evaluation, the number of users is set to 10, 12, 14, and 16. The time interval between each decision making is set to 20ms [14]. In this scenario, the wireless environment changes faster than the resource allocation using the ES, SA, and DQN (train) methods due to their long computation time, as shown in Fig. 9. These methods may no longer apply to mobile NOMA systems. Hence, we only evaluate the performance of the CIM-based and the DQN (test)-based RA methods under these settings. The evaluation results in terms of the ratio of the total data rate and the optimal solution are shown in Fig. 10. From Fig. 10, we can see that the CIM-based RA can achieve a higher data rate than the DQN (test)-based RA regardless of the number of users in the NOMA system. The reason can be summarized as follows. For the DQN (test)-based RA method, the time consumption for resource allocation is longer than the time interval between the decision-making. However, the user location and channel information are changed after the time interval. Hence, the information used by the DQN (test)-based RA method is the previous information that is different from the current information when allocating resources. On the other hand, the time consumption for the CIM-based RA method is shorter than the time interval between the decision-making. The CIM-based RA method can allocate resources using the current information. Hence, the optimal solution of the RA can almost be achieved by the CIM-based RA method. In summary, the proposed CIM-based RA method is applicable to the NOMA system with mobile users.
\begin{figure}[!t]
    \centering
    \includegraphics[width=75mm, clip]{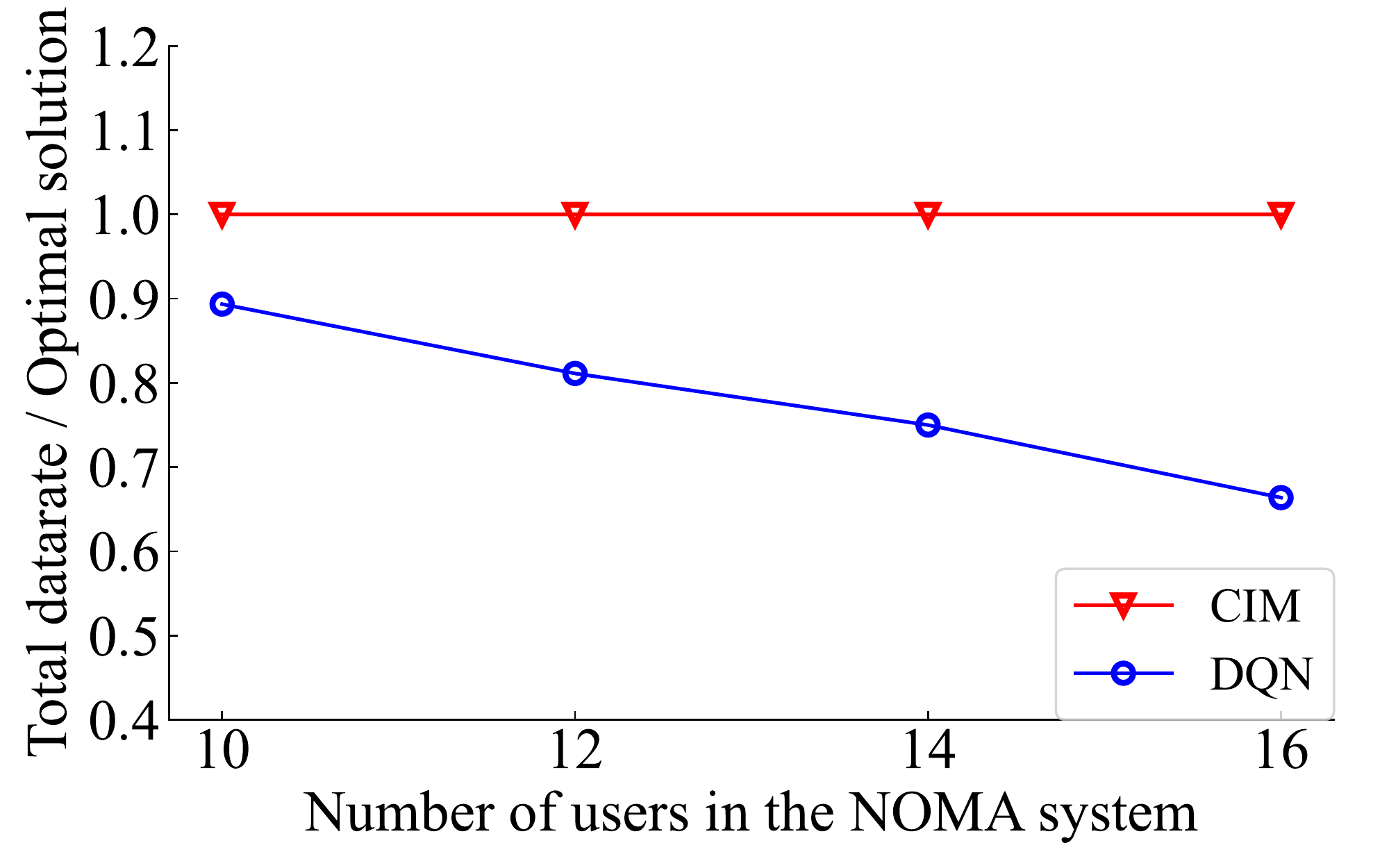}
    \caption{\textcolor{blue}{The ratio of the data rate to the optimal solution in the NOMA system with mobile users.}}
    \label{fig:DQN_CIM}
\end{figure}
\section{Discussion and Future Work}
\label{sect:futurework}

\subsection{Multiple Users Multiplexing in the NOMA System}
In this subsection, we discuss multiplexing more users (more than 2) in the same channel using the proposed CIM-based RA method for the NOMA system. This paper discusses NOMA systems where up to 2 users are multiplexed into the same channel. That is: in Eq. (\ref{eq8}.d), we set a constraint term to limit the maximum number of users that are multiplexing in the same channel to up to 2.
We believe that fast channel allocation in NOMA systems with more multiplexed users can be achieved by extending the constraint term in Eq. (\ref{eq8}.d) as an arbitrary number of users $n$. 
Specifically, given the constraint, the energy function in Eq. (\ref{eq21}) can be calculated. Finally, by recomputing the Ising model parameters, $J_{ijkl}$ and $\lambda_{ij}$, RA in the NOMA system with $n$ users multiplexed can be realized. Note that the number of dummy users $N_d$ to must be set as $N_u+N_d=nN_c$ to satisfy the above constraint. In addition, the number of channel assignment variables $x_{ij}$, $x_{kj}$ in Eq. (\ref{eq8}) needs to be increased to correspond to the number of users multiplexed into the same channel.

\subsection{Scalability of the Proposed Method}
In this subsection, we discuss the scalability of the proposed CIM-based RA methods in the practical settings with 2 or more users in the NOMA systems. As previously described, the CIM is the machine that can obtain the ground state of the Ising Hamiltonian at high speed by getting the optimal combination of the spin directions. The ground state of a quantum-mechanical system is its stationary state of lowest energy, which corresponds to the optimal solution of combinatorial optimization problems. In this paper, the combinatorial optimization problem is the RA problem in the NOMA system. The RA problem in the NOMA system can be solved by transforming the formulated RA combinatorial optimization problem into the form of the Ising Hamiltonian. The spin direction of the spin in the Ising Hamiltonian corresponds to the channel assignment variable of the RA problem in the NOMA system, i.e., whether the channel is allocated to the user or not. Each channel-user pair corresponds to one spin. The value of each spin direction is +1 or -1. If the spin direction is +1, the channel is allocated to the corresponding user. Otherwise, the channel is not allocated to the corresponding user. For example, channel $i$-user $j$ pair corresponding to the spin direction $\sigma_{ij}$. If $\sigma_{ij}$, the channel $i$ is allocated to user $j$. Otherwise, the channel $i$ is not allocated to user $j$. Hence, to solve the RA optimization problem in the NOMA system, $N_u*N_c$ spins are necessary, where $N_u$ and $N_c$ are the numbers of the users and channels, respectively. The related work [19] shows that the CIM can realize problems up to 100000 spins. This means that up to 446 users can be realized using the present CIM in our scenario, where up to 2 users are allocated to one channel simultaneously. In general, however, clustering with an arbitrary number of $n$ users of multiplexing is more attractive. Here, to achieve multiplexing at an arbitrary user $n$ using CIM, we need to change the definition of the assignment variable as described in the last subsection. Specifically, the number of indexes should be increased by considering which user $i$ is multiplexed with which user on channel $j$ in Eq. (\ref{eq9}). Therefore, as n increases, the number of spins required to solve the problem also increases. For example, when $n=3$, the required number of spins is $N_u^3/3$, and RA for up to 67 users can be realized using the present CIM with 100000 spins. With the development of the CIM, much larger-scale optimization problems are expected to be solved by increasing the number of spins in the future.

\subsection{Unequal Bandwidth Allocation}
In this study, the RA in the NOMA system is under the assumption of equal bandwidth allocation. According to \cite{30}, the performance of NOMA systems can be improved by performing unequal bandwidth and allocating higher power and wider bandwidth to users with stronger channel conditions. Therefore, we will consider unequal bandwidth allocation in our future work. We believe that our proposed CIM-based RA method can be easily extended to the NOMA system with unequal bandwidth allocation by assigning more than one resource block to one use, specifically, by considering extending the constraint in Eq. (\ref{eq8}.c) to an arbitrary number n. Using these constraints, the energy function Eq. (\ref{eq20}) can be computed. Finally, by recomputing $J_{ijkl}$ and $\lambda_{ij}$, more than one bandwidth can be allocated to one user using the CIM. Here, by adjusting the value of n, the upper limit of the bandwidth allocated to users with stronger channel conditions can be set. Thus, better system performance may be achieved with unequal bandwidth allocation compared to that with equal bandwidth allocation.

\subsection{Optimal Power Allocation Using the CIM}
In this study, the channel allocation is optimized by the CIM, and then the power is assigned based on the optimizing methods in \cite{21a} using conventional computers. However, in large-scale NOMA systems, the time consumption for allocating power using conventional computers may reduce the speed of the optimization, which may bring a bottleneck to CIM-based optimization. Therefore, channel and power optimization will be jointly considered using the CIM in our future work.

\subsection{Diversity Schemes in NOMA System}
In this study, we assumed that both of the BS and users have a single antenna, and the signal to each user is multiplexed in one channel at the same time. To improve the performance of the system, we will consider diversity schemes such as receiver diversity and antenna diversity in our future work. Specifically, the receiver diversity can improve the communication performance, such as a bit error rate, by combining the same data in the different channels to achieve a diversity gain \cite{35}. Antenna diversity can be used to improve the outage performance of downlink NOMA systems combined with signal user multiple-input multiple-output \cite{36}.

\subsection{Implementation of the CIM}
Although quantum technologies as emerging technologies are expected to be deployed in 6G networks around 2030, it is still unclear when a practical quantum computer will be available. Hence, we will work with the communications and quantum computing industries in the future to try to achieve a practical implementation of the CIM in the next generation of wireless communication.

\section{Conclusion}
\label{sect:conclusion}

In this study, we proposed a CIM-based RA optimization method for NOMA systems.
Initially, we formulated the RA problem as a combinational optimization problem.
To apply RA optimization for NOMA systems in the CIM, we derived the interaction and external magnetic field by mapping the Ising Hamiltonian to the formulated objective function. Herein, the interaction and the external magnetic field are the parameters of the Ising model, which are used to search the optimal solutions.
For correspondence between the Ising Hamiltonian and the formulated objective function, we transformed our formulated objective function into a similar form with an Ising Hamiltonian by the aid of the mutually connected neural network. The parameters of the coupling weights and firing threshold corresponding to the energy function of the mutually connected neural network were then derived.
Subsequently, these parameters were used to derive the Ising model interaction between the spin and external magnetic field used for optimizing the RA in the CIM.
To evaluate the performance of the proposed method, we conducted simulations using the simulation model of the CIM. Moreover, we compared our proposed method with other optimization methods and pairing algorithms. The simulation results indicated that the CIM is not only faster but also superior in searching an optimal solution.

\end{document}